\documentclass[twocolumn]{aastex631} 

\usepackage{amsmath}
\usepackage{xcolor, soul}
\usepackage{ulem}
\usepackage{float}

\newcommand\redout{\bgroup\markoverwith{\textcolor{red}{\rule[0.5ex]{2pt}{0.8pt}}}\ULon}

\hypersetup{linkcolor=cyan,citecolor=teal,filecolor=cyan,urlcolor=blue}

\shorttitle{BH Properties of Type-1 AGNs in the NEP-Wide field}
\shortauthors{Kim et al.}

\begin{document}
\title{Black Hole Properties of Type-1 Active Galactic Nuclei in the North Ecliptic Pole Wide Field: \\
	I. Mid-infrared Sources with Optical Counterparts
	\footnote{myungshin.im@gmail.com (MI); shim.hyunjin@gmail.com (HS)}}

\author[0000-0002-6925-4821]{Dohyeong Kim}
\affiliation{Department of Earth Sciences, Pusan National University, Busan 46241, Republic of Korea}

\author[0000-0002-8537-6714]{Myungshin Im}
\affiliation{SNU Astronomy Research Center (SNUARC), Astronomy Program, Dept. of Physics \& Astronomy, Seoul National University, Seoul 08826, Republic of Korea}

\author[0000-0002-4179-2628]{Hyunjin Shim}
\affiliation{Department of Earth Science Education, Kyungpook National University, 80 Daehak-ro, Buk-gu, Daegu 41566, Republic of Korea}

\author[0000-0002-3560-0781]{Minjin Kim}
\affiliation{Department of Astronomy, Yonsei University, 50 Yonsei-ro, Seodaemun-gu, Seoul 03722, Republic of Korea}
\affiliation{Department of Astronomy and Atmospheric Sciences, College of Natural Sciences, Kyungpook National University, Daegu 41566, Republic of Korea}

\author[0000-0002-5760-8186]{Gu Lim}
\affiliation{Department of Earth Sciences, Pusan National University, Busan 46241, Republic of Korea}

\author[0009-0003-0415-8796]{Junyeong Park}
\affiliation{Department of Earth Sciences, Pusan National University, Busan 46241, Republic of Korea}

\author[0009-0003-4018-9995]{Hayeong Jeong}
\affiliation{Department of Earth Sciences, Pusan National University, Busan 46241, Republic of Korea}

\author[0000-0003-1647-3286]{Yongjung Kim}
\affiliation{School of Liberal Studies, Sejong University, 209 Neungdong-ro, Gwangjin-Gu, Seoul 05006, Republic of Korea}

\author[0000-0003-0134-8968]{Yongmin Yoon}
\affiliation{Department of Astronomy and Atmospheric Sciences, College of Natural Sciences, Kyungpook National University, Daegu 41566, Republic of Korea}

\author[0000-0001-9970-8145]{Seong Jin Kim}
\affiliation{Institute of Astronomy, National Tsing Hua University, 101, Section 2. Kuang-Fu Road, Hsinchu 30013, Taiwan}

\author[0000-0002-3531-7863]{Yoshiki Toba}
\affiliation{Department of Physical Sciences, Ritsumeikan University, 1-1-1 Noji-higashi, Kusatsu, Shiga 525-8577, Japan}
\affiliation{Academia Sinica Institute of Astronomy and Astrophysics, 11F of Astronomy-Mathematics Building, AS/NTU, No.1, Section 4, Roosevelt Road, Taipei 10617, Taiwan}

\author[0000-0002-6821-8669]{Tomotsugu Goto}
\affiliation{Institute of Astronomy, National Tsing Hua University, 101, Section 2. Kuang-Fu Road, Hsinchu 30013, Taiwan}

\author[0000-0002-4686-4985]{Nagisa Oi}
\affiliation{Space Information Center, Hokkaido Information University, Nishi-Nopporo 59-2, Ebetsu, Hokkaido 069-8585, Japan}

\author[0000-0002-4362-4070]{Hyunmi Song}
\affiliation{Department of Astronomy and Space Science, Chungnam National University, 99 Daehak-ro, Yuseong-gu, Daejeon 34134, Republic of Korea}

\begin{abstract}

We present measurements of black hole (BH) properties of
861 Type-1 active galactic nuclei (AGNs) in the North Ecliptic Pole (NEP)-Wide field.
These AGNs are detected in both optical and mid-infrared (MIR) surveys
and are identified as Type-1 AGNs in optical spectroscopic surveys.
By performing spectral energy distribution (SED) and line fitting, 
we obtained their MIR continuum luminosities ($L_{\rm MIR}$)
as well as full width at half maximum (FWHM) values for the \ion{C}{4}, \ion{Mg}{2}, H$\beta$, and H$\alpha$ lines. 
Using these measurements, we derived bolometric luminosities ($10^{43.20}$--$10^{47.27}~{\rm erg~s^{-1}}$),
BH masses ($10^{7.29}$--$10^{9.67}$\,$M_{\odot}$), and Eddington ratios ($10^{-2.74}$--$10^{-0.08}$)
for $\sim$450 objects over a wide redshift range ($z=0.09$--$4.71$).
The use of $L_{\rm MIR}$ and FWHM values effectively alleviates the effects of dust extinction, 
enabling reliable estimates of BH properties even for dust-obscured AGNs.
Moreover, we find that 34\,\% of the Type-1 AGNs in the NEP-Wide field are dust-obscured, 
and that their bolometric luminosities can be significantly underestimated without proper dust extinction correction.
Our relatively extinction-free BH property estimates can (i) be combined with multi-wavelength data in the NEP-Wide field
to facilitate diverse studies of AGN environments, number densities, host galaxies, and related topics,
and (ii) serve as fiducial estimates for SPHEREx and other upcoming infrared (IR) spectroscopic missions covering the NEP-Wide field.

\end{abstract}


\section{Introduction} \label{sec:intro}
After the launch of the $\it{AKARI}$ \citep{murakami07} IR space telescope in February 2006,
a targeted survey of the NEP
(R.A.\ \(=\) $18^{\rm h}00^{\rm m}00^{\rm s}$, \quad Dec.\ \(=\) $+66^\circ33'88''$) was carried out.
The NEP field was strategically selected owing to its high visibility from $\it{AKARI}$'s polar Sun-synchronous orbit, 
making it optimal for repeated observations.
The $\it{AKARI}$ NEP survey \citep{matsuhara06} consists of two complementary programs
-- NEP-Deep and NEP-Wide -- conducted with InfraRed Camera (IRC; \citealt{onaka07})
using infrared filter sets from 2\,$\mu$m to 24\,$\mu$m.
Of these, the NEP-Wide survey covers an area of 5.4\,$\rm{deg^{2}}$,
with exposure times ranging from 90 to 150\,seconds \citep{lee09,kim12}.

The $\it{AKARI}$ NEP-Wide survey is notably complemented by 
extensive multi-wavelength photometric data from various ground-based observations. 
Optical photometric data \citep{hwang07,jeon10} were provided by MegaCam on the Canada-France-Hawaii Telescope (CFHT) 
and by the Seoul National University Camera (SNUCAM; \citealt{im10}) at Maidanak Observatory. 
More recently, deeper and more comprehensive optical photometry \citep{oi21} covering the entire NEP-Wide field was obtained 
using the Hyper Suprime-Cam (HSC; \citealt{miyazaki18}) on the Subaru Telescope. 
Additionally, near-infrared (NIR) $\it{J}$- and $\it{H}$-band photometric data \citep{jeon14} were acquired with 
FLAMINGOS at the Kitt Peak National Observatory (KPNO), 
and 850\,$\mu$m sub-mm observations \citep{shim20} were conducted using 
the Submillimetre Common-User Bolometric Array 2 (SCUBA-2) on the James Clerk Maxwell Telescope (JCMT).
Furthermore, the NEP-Wide field was also observed at 20\,cm \citep{white10} over an area of 1.7\,$\rm{deg^{2}}$ using
the Westerbork Synthesis Radio Telescope (WSRT).

Given its role as a preferred deep-field region for astronomical satellites,
the NEP-Wide field has been extensively observed, both photometrically and spectroscopically,
by various space telescopes such as SPHEREx \citep{dore16,dore18}, $\it{Euclid}$ \citep{euclid22,euclid25}, 
eROSITA \citep{merloni12,predehi21}, $\it{ROSAT}$ \citep{truemper82,henry06}, 
$\it{GALEX}$ \citep{martin05,burgarella19}, 
$\it{Herschel}$ \citep{pilbratt10,burgarella19,pearson19},
WISE \citep{wright10,jarrett11}, and $\it{Spitzer}$ \citep{werner04,jarrett11,nayyeri18}.
It is worth noting that, among these missions, SPHEREx is uniquely positioned to provide IR spectroscopic data for all objects in the NEP-Wide field,
offering new insights that cannot be derived from photometric observations alone. 

Although the NEP-Wide field is well-covered with various photometric and spectroscopic surveys from X-ray to radio,
optical spectroscopic observations remain relatively sparse.
\cite{shim13} provided optical spectra of 1796 objects selected in the NEP-Wide field,
which were obtained with Hectospec on the MMT \citep{fabricant08} and 
Hydra on the WIYN telescope at KPNO \citep{barden94}.
Using these spectra, they identified 1128 galaxies, 198 type-1 AGNs, and 121 stars.
Furthermore, the Dark Energy Spectroscopic Instrument (DESI; \citealt{desi16a,desi16b}) collaboration
recently released Data Release 1 (DR1; \citealt{desi25}) data, which fully covers the NEP-Wide field.
The DESI DR1 data include optical spectra for 18 million objects,
among which 13 million are classified as galaxies, 1.6 million as quasars, and 4 million as stars.

AGNs are generally classified into Type-1 and Type-2 based on their spectroscopic properties.
In the AGN unification model \citep{urry95}, this classification primarily reflects
the orientation of the dust torus surrounding the central SMBH and its accretion disk relative to the observer's line of sight.
Type-1 AGNs show both broad and narrow emission lines (BELs and NELs),
whereas Type-2 AGNs exhibit only NELs due to obscuration of the region emitting BELs by the dust torus.
Beyond the orientation-based dichotomy, there exists another population of dust-obscured AGNs,
which are thought to be affected by dust extinction originating in their host galaxies rather than in the dust torus (e.g., \citealt{mezcua16}).
They exhibit BELs similar to those of Type-1 AGNs but have very large Balmer decrements,
or exhibit only NELs at short wavelengths like Type-2 AGNs while exhibiting BELs at longer wavelengths. 

To understand fundamental properties of AGNs, optical spectroscopic data are required.
In particular, BH masses (hereafter, $M_{\rm BH}$) can be measured from optical spectroscopy,
using the velocity widths of BELs and the sizes of broad-line regions (BLRs).
The velocity widths of BELs are commonly expressed as the full width at half maxima (FWHMs) of the lines.
While the sizes of the BLRs can be directly measured using the reverberation mapping method \citep{peterson04},
this method is observationally intensive and time-consuming.
Hence, the BLR sizes are more commonly measured from single-epoch spectra,
using empirical relations between UV or optical continuum luminosities and the BLR sizes (e.g., \citealt{mclure04,kaspi05}). 

However, the optical spectra easily suffer significantly from dust extinction (e.g., \citealt{kim18a,kim25}).
For a dust-obscured AGN with a color excess of $E(B-V)=2$, the continuum luminosity at 5100\,$\rm \AA{}$ (hereafter, L5100) decreases by a factor of 500,
but the P$\alpha$ (1.87\,$\mu$m) and Br$\alpha$ (4.05\,$\mu$m) lines decrease only by factors of 2.16 and 1.31, respectively.
Hence, alternative approaches using IR spectroscopic data have been developed to estimate BH masses of dust-obscured AGNs.
For example, Hydrogen Paschen lines \citep{kim10,landt13,franca15,ricci17,kim22} and Brackett lines \citep{kim15b,kim22}
were newly established as BH mass estimators.

Nevertheless, due to the limited availability of IR spectroscopic data compared to optical spectroscopy,
applying these Paschen and Brackett line-based BH mass estimators remains challenging.
To address this limitation, \cite{kim23} established new BH mass estimators
based on $L_{\rm MIR}$ in combination with optical spectroscopy.
In this approach, optical spectra provide the velocity width of the BELs,
while the $L_{\rm MIR}$ values serve as proxies for the BLR sizes.
These $L_{\rm MIR}$ values are derived from multi-wavelength photometry ranging from optical to MIR,
and the resulting $L_{\rm MIR}$-based estimators \citep{kim23} are relatively free from dust extinction.
However, it is worth noting that the IR luminosity may be affected by contamination from the host galaxy. 
In particular, the host galaxy contribution can be significant ($\sim$50\,\%) in the NIR,
somewhat less so but non-negligible ($\lesssim 20\,\%$) for some AGNs in the MIR \citep{kim23}.
Nevertheless, these $L_{\rm MIR}$-based estimators have been successfully applied to 
both unobscured and dust-obscured AGNs through SED fitting that includes host galaxy component
\citep{kim24a,kim24b,kim25}.

The aim of this work is to present BH properties of Type-1 AGNs in the NEP-Wide field
detected in both optical and MIR surveys. 
This wide wavelength coverage allows us to apply the $L_{\rm MIR}$-based estimators \citep{kim23}
to derive reliable BH properties that are less affected by dust extinction.
Combining these BH properties with existing multi-wavelength observations in the NEP-Wide field
could facilitate future studies exploring how various factors,
such as environment and host galaxy morphology, affect the BH properties. 
Furthermore, with the upcoming IR low-resolution spectroscopic survey by SPHEREx covering the NEP-Wide field,
our results may provide fiducial values for understanding how the IR low-resolution spectroscopy
affects the measurements of the fundamental BH properties. 
Throughout this work, we adopt a standard $\Lambda$CDM model of 
$H_{\rm 0}=70\,{\rm km\,s^{-1}}$\,Mpc$^{-1}$, $\Omega_{\rm m}=0.3$, and $\Omega_{\rm \Lambda}=0.7$,
which is supported by several observational studies (e.g., \citealt{im97,planck16}).
All magnitudes are given in the AB magnitude system.  

\section{Sample} \label{sec:sample}

Our sample is based on the MIR source catalog in the NEP-Wide field \citep{kim21},
which was constructed by cross-matching MIR sources \citep{kim12} with a deep optical catalog \citep{oi21}.
The MIR observations were carried out by the IRC \citep{onaka07} aboard the $\it{AKARI}$ space telescope,
and details of the data reduction, detection, and photometry are provided in \cite{kim12}.
Each source has photometric measurements in nine bands obtained with the IRC/$\it{AKARI}$,
which provide continuous coverage from 2\,$\mu$m to 25\,$\mu$m,
including N2, N3, and N4 (NIR); S7, S9W, and S11 (MIR-S); and L15, L18W, and L24 (MIR-L).

Moreover, the NEP-Wide field was covered by a deep optical survey \citep{oi21} using the HSC \citep{miyazaki18} on the Subaru telescope,
with five photometric bands: $g$, $r$, $i$, $z$, and $y$.
This photometric catalog of the HSC survey reaches a 5$\sigma$ depth of $\sim$28.1\,mag in the $\it{g}$ band,
which is $\sim$2\,mag deeper than previous optical surveys in the NEP-Wide field (e.g., \citealt{hwang07}).

\cite{kim21} present 91,861 $\it{AKARI}$ MIR sources that were cross-matched with the HSC optical catalog
using a matching radius of 1.78\,arcseconds, of which 95\,\% were matched within 1.5\,arcseconds.
The matching radius was determined as the 3$\sigma$ value from the positional offset distributions of the matched sources.
These matched sources (hereafter, $\it{AKARI}$--HSC sources) serve as the parent sample
from which our sample of Type-1 AGNs is drawn.
Besides the $\it{AKARI}$--HSC sources, 43,294 MIR sources were excluded,
which were either not matched to any optical sources
or were flagged as unreliable matches due to being located near masked edges or affected by saturated pixels.
The detailed description of the matching process is provided in \cite{kim21}.
Consequently, the $\it{AKARI}$--HSC sources exclude MIR sources that are not detected in the optical survey,
which are considered to be either (i) obscured objects or (ii) high-$\it{z}$ sources \citep{toba20}.

In addition, \cite{kim21} matched various photometric catalogs to the $\it{AKARI}$--HSC sources,
providing a complementary photometric data set.
This complementary data set includes
optical data from MegaPrime/CFHT \citep{huang20}, MegaCam/CFHT \citep{hwang07}, and SNUCAM/Maidanak \citep{jeon10};
IR data from FLAMINGOS/KPNO \citep{jeon14}, IRAC/$\it{Spitzer}$ \citep{nayyeri18}, WISE \citep{jarrett11},
PACS/$\it{Herschel}$ \citep{pearson19}, and SPIRE/$\it{Herschel}$ \citep{pearson17};
and sub-mm measurements from SCUBA-2/JCMT \citep{shim20}.

We matched the $\it{AKARI}$--HSC sources with the Type-1 AGNs identified from two different optical spectroscopic surveys.
First, we used the quasars from the DESI DR1 \citep{desi25}.
In the DESI DR1, quasar candidates were selected using three optical bands ($\it g$, $\it r$, and $\it z$) and two IR bands (W1 and W2; \citealt{chaussidon23}).
These candidates were spectroscopically confirmed as quasars using
the DESI pipeline classifier \texttt{Redrock} (Bailey et al. in prep.),
the machine-learning algorithm \texttt{QuasarNet} \citep{busca18},
and the \ion{Mg}{2} post-processing pipeline, 
which further improves quasar classification and redshift estimation \citep{chaussidon23,alexander23}.
Note that the DESI quasar selection is optimized for point-like sources and
is effectively limited to 16.5\,mag $\lesssim \it{r} \lesssim$ 23\,mag \citep{chaussidon23}.
Therefore, this sample may miss Type-1 AGNs that are brighter than the bright limit ($\it{r} \lesssim$ 16.5\,mag),
fainter than the faint limit ($\it{r} \gtrsim$ 23\,mag),
or whose photometry is significantly contaminated by host galaxy light.

We matched the DESI DR1 catalog to the $\it{AKARI}$--HSC sources using a matching radius of 2.2\,arcseconds,
which corresponds to twice the seeing size in the $\it{z}$ band of the DESI Legacy Imaging Survey \citep{dey19}.
This yields 14,797 matched sources, of which 832 are classified as DESI quasars.
Note that the false-match rate is expected to be $\sim$2.2\,\%.
To estimate the false-match rate, we performed a control test by randomly shifting the DESI quasar positions by 10\,arcseconds
and repeating the cross-matching with the $\it{AKARI}$--HSC sources using the same 2.2\,arcsecond radius.
In total, 831 DESI DR1 quasars were cross-matched with the $\it{AKARI}$--HSC sources.
Note that, among the 831 DESI DR1 quasars, one object was matched with two $\it{AKARI}$--HSC sources,
and the nearest $\it{AKARI}$--HSC source was adopted as the counterpart.

Second, we used the Type-1 AGNs identified by \cite{shim13},
who selected Type-1 AGN candidates based on two IR color criteria, $\rm N2 - N4 > 0$ and $\rm S7 - S11 > 0$.
Using Hectospec on the MMT and Hydra on the WIYN, 
they obtained optical spectra of the AGN candidates with $\it{R}$-band magnitudes brighter than 22.5,
among which 198 were identified as Type-1 AGNs. 
Among them, we found 163 Type-1 AGNs that were matched to the $\it{AKARI}$--HSC sources.
However, since 133 of them were already identified as DESI quasars,
we used only the remaining 30 objects as Type-1 AGNs from \cite{shim13}.

\begin{figure}
\figurenum{1}
\includegraphics[width=\columnwidth]{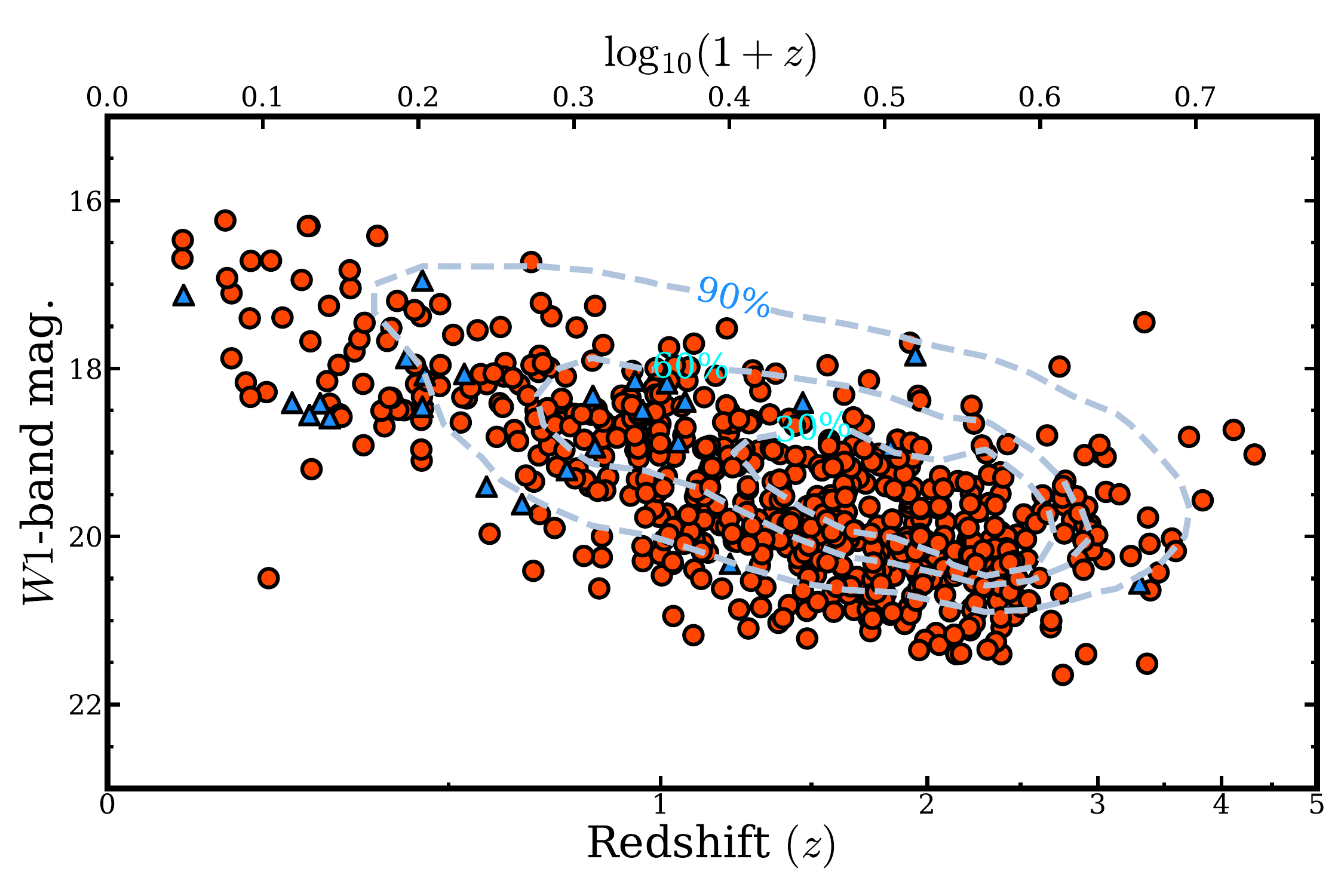}
\caption{
Redshift versus $\it{W1}$-band magnitude. 
Red circles and blue triangles represent the 861 Type-1 AGNs used in this study,		
identified from DESI DR1 and \cite{shim13}, respectively.
The gray dashed contours present the distribution of redshifts and $\it{W1}$-band magnitudes of SDSS DR14 quasars for comparison.
\label{fig:Sample}}
\end{figure}

In conclusion, a total of 861 $\it{AKARI}$--HSC sources were identified as Type-1 AGNs,
which were used as the main sample in this study. 
In this paper, the term ``AGN" encompasses both luminous quasars and lower-luminosity AGNs.
Figure \ref{fig:Sample} presents the distributions of $\it{W1}$-band magnitudes and redshifts for our sample,
which are not significantly different from those of Sloan Digital Sky Survey (SDSS) Data Release 14 (DR14) quasars \citep{paris18,rakshit20}.
Our sample spans a wide range of redshifts ($0.09 < z < 4.71$), BH mass ($10^{7.29} < M_{\rm BH} / M_{\odot} < 10^{9.67} $), 
and bolometric luminosity (hereafter, $L_{\rm bol}$; $10^{43.20} < L_{\rm bol} / {\rm erg~s^{-1}} < 10^{47.27}$).
Note that these values were estimated using the $L_{\rm MIR}$-based estimators,
as described in Sections \ref{sec:Lbol} and \ref{sec:MBH}.

	\begin{deluxetable*}{cccccccccccc}
		\tabletypesize{\tiny}
		\tablecolumns{12}
		\tablewidth{0pt}
		\tablenum{1}
		\tablecaption{Basic Properties and Measurements of Type-1 AGNs in the NEP-Wide Field\label{tbl:info_est}}
		\tablehead{
			\colhead{$\it AKARI$ ID} & \colhead{Spec. ID\tablenotemark{a}} & \colhead{$\alpha_{\rm J2000}$} & \colhead{$\delta_{\rm J2000}$} & \colhead{redshift} &
			\colhead{$\log \left( \frac{L_{\rm 3.4}}{\rm erg~s^{-1}} \right)$\tablenotemark{b}} & 
			\colhead{$\log \left( \frac{L_{\rm 4.6}}{\rm erg~s^{-1}} \right)$\tablenotemark{b}} &
			\colhead{$E(B-V)$}&	\multicolumn{4}{c}{FWHM}\\
			\cline{9-12}
			\colhead{}& \colhead{}& \colhead{}& \colhead{}& \colhead{}& 
			\colhead{}& \colhead{}& \colhead{}&
			\colhead{\ion{C}{4}} & \colhead{\ion{Mg}{2}} & \colhead{H$\beta$} & \colhead{H$\alpha$} \\ [-8pt]
			\colhead{}& \colhead{}& \colhead{(deg.)}& \colhead{(deg.)}& \colhead{($\it z$)}& 
			\colhead{}& \colhead{}& \colhead{(mag.)}&
			\colhead{($\rm km~s^{-1}$)}& \colhead{($\rm km~s^{-1}$)}& \colhead{($\rm km~s^{-1}$)}& \colhead{($\rm km~s^{-1}$)}
		}
		\startdata
		133109 &  39633470514793582  &  269.5586  &  67.7215 &  0.205 &
		$43.82 \pm 0.02$  &  $43.81\pm0.02$ &  0.00  &
		--  &  -- & $3760\pm120$ & $2840\pm30$ \\
		135036  &  39633461161495358  &  272.9552  &  66.7585 &  0.996 &
		$44.71 \pm 0.02$  &  $44.69 \pm 0.02$&  0.00  &
		--  &  $3840 \pm 710$  &  --  &  --\\
		44872  &  39633456321267046  &  272.0693  &  66.1369 &   0.293 &
		-- &  $42.96 \pm 0.03$ &  0.00  &
		--  &  --  &  --  &  $5290 \pm 200$ \\
		134539  &  39633463535471858  &  272.0972  &  67.0477  &  0.250 &
		--  &  --  &  --  &
		--  &  --  &  --  &  $9360 \pm 1020$\\
		133581  &  39633465880087464  &  270.4529  &  67.3362  &  0.333  &
		$43.81 \pm 0.02$  &  $43.79 \pm 0.02$  &  0.00  &
		--  &  --  &  $3150 \pm 120$  &  $2750 \pm 60$\\
		46186  &  2305843019946926614  &  269.2281  &  65.9080  &  0.509  &
		$44.18 \pm 0.02$  &  $44.16 \pm 0.02$  &  0.00  &
		--  &  --  &  --  &  --  \\ 
		133141  &  39633448859602050  &  269.6010  &  65.5188  &  0.326  & 
		$44.10 \pm 0.02$  &  $44.08 \pm 0.02$  &  0.00  &
		--  &  $3200 \pm 290$  &  $3990 \pm 180$  &  $3740 \pm 290$ \\
		132335  &  39633448851210769  &  267.9706  &  65.4902  &  0.401  & 
		--  &  --  &  --  &
		--  &  --  &  --  &  -- \\
		132310  &  39633451355212011  &  267.9039  &  65.6752  &  0.825  & 
		$45.02 \pm 0.02$  &  $45.00 \pm 0.02$  &  0.22  &
		--  &  $3260 \pm 220$  &  $3610 \pm 120$  &  -- \\
		132768  &  39633453842434564  &  268.8884  &  66.116  &  0.198  & 
		--  &  $42.84 \pm 0.03$  &  0.00  &
		--  &  --  &  --  &  -- \\
		133094  &  39633453846628993  &  269.5275  &  66.0083  &  0.087  & 
		--  &  --  &  --  &
		--  &  --  &  --  &  $5560 \pm 220$ \\
		133299  &  39633453850822123  &  269.910  &  65.9765  &  2.235  & 
		$45.57 \pm 0.03$  &  $45.55 \pm 0.03$  &  0.00  &
		--  &  --  &  --  &  -- \\
		132336  &  39633453838238503  &  267.9721  &  65.9240  &  0.307  & 
		--  &  $43.28 \pm 0.03$  &  0.00  &
		--  &  --  &  --  &  -- \\
		53113  &  39633453838239265  &  268.0851  &  65.9499  &  0.348  & 
		--  &  --  &  --  &
		--  &  --  &  --  &  -- \\
		132128  &  39633456287715657  &  267.5017  &  66.2005  &  0.524  & 
		--  &  --  &   --  &
		--  &  $6050 \pm 1200$  &  --  &  -- \\
		132658  &  39633458728797750  &  268.6906  &  66.3971  &  0.088  & 
		--  &  --  &  --  &
		--  &  --  &  $4870 \pm 500$  &  $2930 \pm 40$ \\
		132883  &  39633458732990946  &  269.1327  &  66.4189  &  0.263  & 
		$43.59 \pm 0.02$  &  $43.57 \pm 0.02$  &  0.28  &
		--  &  --  &  --  &  -- \\
		\enddata
		\tablenotetext{a}{Spectroscopic ID adopted from DESI DR1 and \cite{shim13}.}
		\tablenotetext{b}{Extinction-corrected $L_{\rm MIR}$ values.}
		\tablecomments{Columns (1)–(2): $\it AKARI$ and spectroscopic ID. Columns (3)–(4): Right Ascension and Declination in degrees. Column (5): Redshift. 
			Columns (6)–(7): $L_{\rm MIR}$ values. Column (8): Dust extinction $E(B-V)$. Columns (9)–(12): FWHMs of emission lines.
			This table is fully available in a machine-readable format.}
	\end{deluxetable*}

\section{Analysis} \label{sec:analysis}
\subsection{SED fitting} \label{sec:SED}

In order to estimate $L_{\rm bol}$ and $M_{\rm BH}$ using the $L_{\rm MIR}$-based estimators \citep{kim23},
measurements of the monochromatic continuum luminosities, $\lambda L_{\lambda}$, 
at rest-frame 3.4\,$\mu$m and 4.6\,$\mu$m (hereafter, $L_{\rm 3.4}$ and $L_{\rm 4.6}$, respectively) are required.
To obtain the $L_{\rm 3.4}$ and $L_{\rm 4.6}$ values for our sample,
we performed SED fitting using multi-wavelength photometry ranging from optical to MIR.

For the SED fitting, we used the photometric data from
the HSC/Subaru ($\it{g}$, $\it{r}$, $\it{i}$, $\it{z}$, and $\it{y}$ bands; \citealt{oi21}),
the FLAMINGOS/KPNO ($\it{J}$ and $\it{H}$ bands; \citealt{jeon14}),
the IRC/$\it{AKARI}$ (N2, N3, N4, S7, S9W, S11, L15, L18W, and L24 bands; \citealt{kim12}),
the WISE ($\it{W1}$, $\it{W2}$, and $\it{W3}$ bands; \citealt{jarrett11}),
and the IRAC/$\it{Spitzer}$ (IRAC1, IRAC2, IRAC3, and IRAC4 bands; \citealt{nayyeri18}).
These photometric data, originally obtained from the multiple surveys,
were cross-matched and assembled into a unified catalog by \cite{kim21},
from which we adopted the photometry for our analysis.
An intrinsic fluctuation of $\sigma_{m}$=0.035\,mag \citep{kim15b}, arising from AGN variability,
was added in quadrature to the original photometric uncertainties in each band.
In addition, only photometric data with signal-to-noise ratios (S/N) greater than 3 were included in the SED fitting.
Note that the WISE $\it{W4}$ photometry was excluded from the SED fitting
owing to its photometric calibration issues \citep{brown14}.

The SED fitting was performed following the methodology described in \cite{kim23}.
The photometric flux density, $f(\lambda)$, is fitted with a sum of reddened spectra:
\begin{equation}
	f(\lambda) = C_{1} A(\lambda) + C_{2} E(\lambda) + C_{3} S(\lambda)  + C_{4} I(\lambda), \label{eqn:SED}
\end{equation}
where $A(\lambda)$, $E(\lambda)$, $S(\lambda)$, and $I(\lambda)$
represent reddened AGN, elliptical, spiral, and irregular galaxy spectra, respectively,
with each component scaled by its corresponding coefficient, denoted $C_{1}$ through $C_{4}$.
These reddened AGN and galaxy spectra are derived from
the intrinsic spectral templates $A_{0}(\lambda)$, $E_{0}(\lambda)$, $S_{0}(\lambda)$, and $I_{0}(\lambda)$.
The intrinsic AGN template is adopted from \cite{krawczyk13},
whereas the galaxy templates are taken from \cite{assef10}.
While several alternative AGN templates are available (e.g., \citealt{richards06,assef10}),
they introduce no significant difference in the SED fitting results \citep{kim23,kim24b,kim25}.

For the SED fitting, we used the $f(\lambda)$ that had already been corrected for Galactic extinction.
The extinction corrections for each band were adopted from \cite{oi21},
based on a reddening value of $E(B-V) = 0.046$ for the NEP-Wide field \citep{schlegel98}.
In addition, intrinsic extinction (i.e., dust obscuration within the host galaxy) was applied to 
the spectral templates to obtain the reddened spectra.
To account for dust extinction, we adopted the reddening law of \cite{cardelli89}, assuming $\it{R_{V}}$=3.1 (e.g., \citealt{weingartner01}).
Since this law is significantly different from other reddening laws (e.g., \citealt{calzetti00}) in the UV due to the 2175\,$\rm \AA{}$ bump,
we excluded the $f(\lambda)$ values at wavelengths shorter than 3000\,$\rm \AA{}$ from the SED fitting.

Note that the NEP-Wide field covers a large area,
while the spatial variations in the reddening values are small.
The measured $E(B-V)$ values in this field \citep{schlegel98} range from 0.044 to 0.047.
Moreover, the reddening value can be estimated differently
depending on the attenuation curve or adopted SED model (e.g., \citealt{galametz17}).
For instance, \cite{schlafly11} reported $E(B-V)=0.040$ for the NEP-Wide field.
However, such differences have a negligible effect on the SED fitting,
producing an extinction-corrected magnitude difference of only $\sim$0.03\,mag even at 3000\,$\rm \AA{}$,
which is the shortest fitting wavelength used in the SED fitting.

\begin{figure*}
\centering
\figurenum{2}
\includegraphics[width=\textwidth]{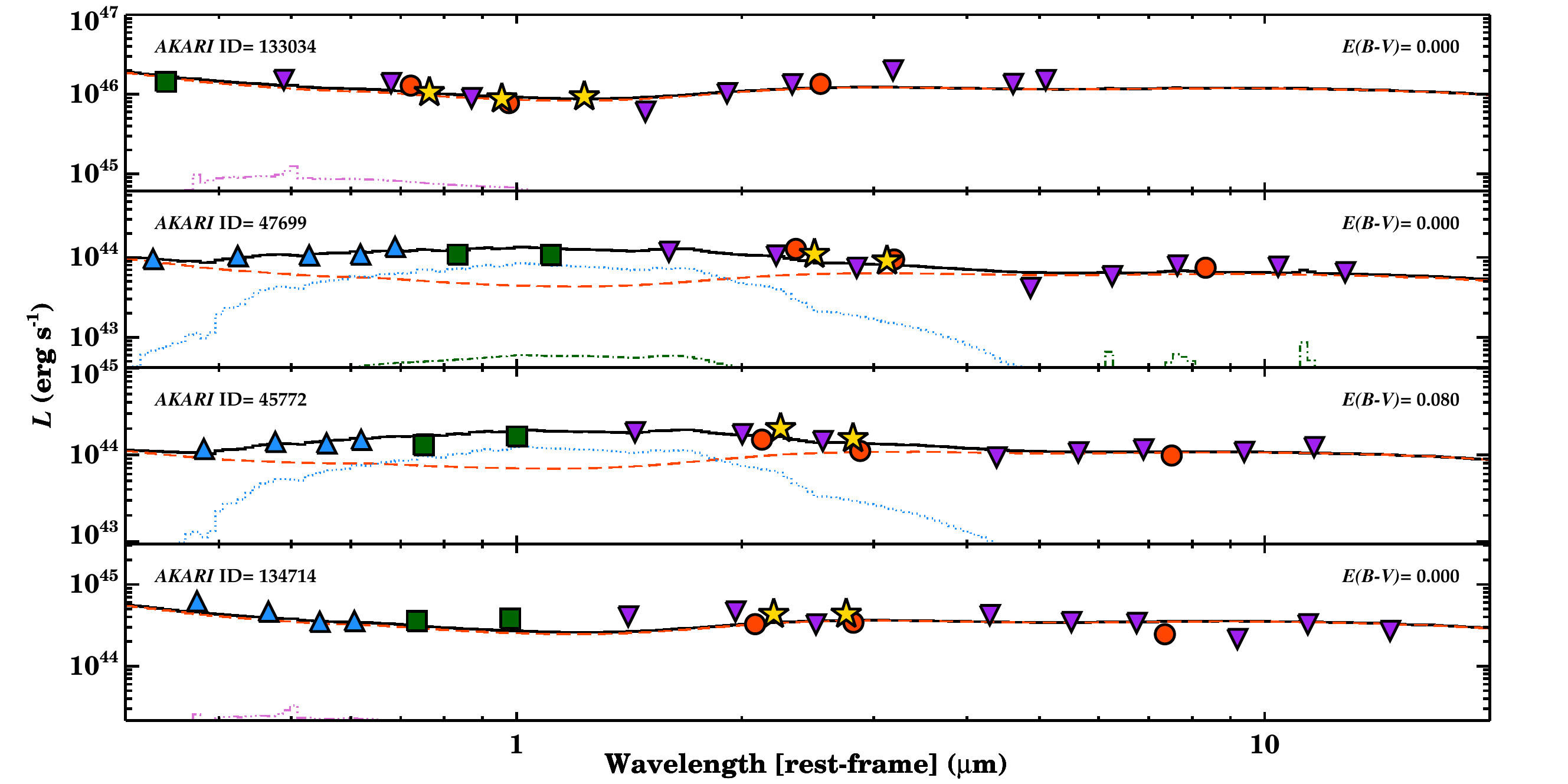}
\caption{
Best-fit SED results and corresponding photometric data for four randomly selected Type-1 AGNs.
Blue triangles, green squares, purple triangles, red circles, and yellow stars represent
the photometric data from HSC/Subaru \citep{oi21}, FLAMINGOS/KPNO \citep{jeon14},
IRC/$\it{AKARI}$ \citep{kim12}, WISE \citep{jarrett11}, and IRAC/$\it{Spitzer}$ \citep{nayyeri18}, respectively.
The solid black line shows the best-fit SED model, 
while the dashed red, dotted blue, dot-dashed green, and dot-dot-dashed purple lines represent 
the reddened spectra of AGN, elliptical, spiral, and irregular galaxies, respectively.
\label{fig:SED}}
\end{figure*}

We used the \texttt{MPFIT} \citep{markwardt09} procedure in IDL for the SED fitting,
and representative examples of the fitting results are shown in Figure \ref{fig:SED}.
From the SED fitting, we obtained the $L_{\rm 3.4}$, $L_{\rm 4.6}$, and $E(B-V)$ values with their uncertainties for our sample,
and these uncertainties are the formal errors derived from the fitting routine based on the input flux uncertainties.
Note that we performed the SED fitting only when the number of photometric data exceeded five.

However, we only used these estimates if the following criteria were met.
For the $L_{\rm 3.4}$, the AGN fraction at 3.4\,$\mu$m had to be greater than 0.5
(i.e., $C_{1}A({\rm 3.4\,\mu m}) >  \frac{1}{2} f({\rm 3.4\,\mu m})$ in Equation \ref{eqn:SED}),
and the SED fitting had to include at least one photometric data point at rest-frame wavelength longer than 3.4\,$\mu$m.
Similarly, the $L_{\rm 4.6}$ values required an AGN fraction greater than 0.5 at 4.6\,$\mu$m,
and at least one photometric data point beyond 4.6\,$\mu$m in the SED fitting.
Note that, when the AGN fraction is low, the MIR SED can be dominated by the host galaxy,
which may compromise the reliability of the $L_{\rm MIR}$ measurement.
Moreover, in both cases, we required at least one photometric data point at rest-frame wavelengths shorter than 0.8\,$\mu$m,
since reliable $E(B-V)$ estimates cannot be obtained in the absence of short-wavelength data.
At $z \gtrsim 2.4$, this becomes challenging because
the HSC photometric bands correspond to rest-frame wavelengths below 0.3\,$\mu$m
and were thus excluded from the SED fitting,
while the relatively shallow NIR photometry often fails to provide sufficient S/N data points.

Finally, using the reliable $L_{\rm 3.4}$, $L_{\rm 4.6}$, and $E(B-V)$ values,
we obtained extinction-corrected $L_{\rm 3.4}$ and $L_{\rm 4.6}$ values for 609 and 532 objects, respectively.
These extinction-corrected luminosities are listed in Table \ref{tbl:info_est}.
Note that the $E(B-V)$ estimates could be affected by various intrinsic AGN continuum slopes,
but the effects of such continuum slope variations on the $L_{\rm MIR}$ measurements are not significant \citep{ykim24}.

\subsection{Line fitting} \label{sec:line}

In order to estimate $M_{\rm BH}$ using the $L_{\rm MIR}$-based estimators \citep{kim23},
we fitted the H$\alpha$, H$\beta$, \ion{Mg}{2}, and \ion{C}{4} lines to obtain the FWHMs of their BEL components.
For this line fitting, we used the optical spectra from DESI DR1 \citep{desi25} and \cite{shim13}
after applying Galactic extinction corrections as described in Section \ref{sec:SED}.
Note that the DESI spectra cover wavelengths from 3600 to 9800\,$\rm \AA{}$.
\cite{shim13} provide spectroscopic data obtained with Hectospec/MMT and Hydra/WIYN,
which cover 3700--8500\,$\rm \AA{}$ and 4500--9000\,$\rm \AA{}$, respectively.
The wavelength ranges of these datasets are broadly comparable.

To fit these lines, we used the \texttt{PyQSOFit} code \citep{guo18,shen19} in Python.
For each line, the fitting was performed separately using the spectral region within $\pm 1000\,{\rm \AA{}}$ of the line center.
During the fitting process, the spectra were simultaneously decomposed into continuum and emission-line components,
with the continuum modeled as a combination of a single power-law and
an empirical \ion{Fe}{2} emission template \citep{shen19}.
Note that host galaxy subtraction was not performed during the continuum fitting.
However, since the continuum fitting was restricted to a relatively narrow wavelength range,
the host galaxy contribution was effectively incorporated into the power-law model,
and the FWHM measurements of the BELs are not significantly affected.

After subtracting the continuum component, we simultaneously fitted the emission line components, 
which consist of multiple line complexes.
The \ion{Mg}{2} complex includes both broad and narrow \ion{Mg}{2} lines.
The H$\beta$ complex comprises broad and narrow H$\beta$ lines,
as well as the [\ion{O}{3}] $\lambda \lambda$4959, 5007 doublet (including core and wing components).
The H$\alpha$ complex includes broad and narrow H$\alpha$ lines,
along with the [\ion{N}{2}] $\lambda \lambda$6549, 6585 and [\ion{S}{2}] $\lambda \lambda$6718, 6732 doublets.
In contrast, the \ion{C}{4} line is modeled with only a broad line, as it does not form a line complex. 
Note that the [\ion{O}{3}] line often exhibits a wing component (e.g., \citealt{singha22}),
which is commonly interpreted as evidence of gas outflows.
A single Gaussian (representing only the core component) fit to the [\ion{O}{3}] profile
can therefore yield unreliable estimates of the line width and velocity offset,
and we find that the H$\beta$ FWHM estimate can differ by $\sim$3.5\,\% when the [\ion{O}{3}] wing component is omitted.

To fit the broad lines, we used multi-Gaussian models:
three Gaussians were used for \ion{C}{4}, H$\beta$, and H$\alpha$, and two for \ion{Mg}{2}.
These Gaussian models follow the fitting scheme adopted by \cite{rakshit20}.
To obtain the FWHM values for the broad lines,
we measured the FWHM from the composite profile of the multi-Gaussian model,
following widely adopted methods in previous studies (e.g., \citealt{rakshit20,shen19}).
Occasionally, an additional, extremely broad and faint Gaussian component was included,
but it contributes negligibly to the composite profile and does not affect the derived FWHM value.
Note that the uncertainties in the FWHM estimates were derived using a Monte Carlo simulation.
Each Gaussian component was randomly varied with its 1\,$\sigma$ error,
and the FWHM was recalculated over 1000 random samples.
The standard deviation of the resulting FWHM distribution was adopted as the FWHM uncertainty.

In this fitting, the broad lines were defined as
those having FWHM values exceeding 2000\,$\rm km~s^{-1}$, following \cite{kim23,kim24b}.
Although various criteria for broad lines have been adopted in other studies (e.g., \citealt{xu07,suh19}),
this criterion effectively excludes narrow-line Seyfert 1 galaxies  (e.g., \citealt{zhou06}). 
Moreover, the narrow lines were constrained to have FWHM values less than 1000\,$\rm km~s^{-1}$.
For these lines, the core components were required to have velocity offsets within $\pm$1000\,$\rm km~s^{-1}$,
while the wing components were allowed to shift up to 3000\,$\rm km~s^{-1}$ from the line center.

Note that the \ion{C}{4} line was not fitted when strong broad absorption features were present,
such as broad absorption lines (BALs), which are believed to originate from high-velocity outflowing gas.
In such cases, both the continuum and the emission-line components were excluded from fitting,
since reliable FWHM measurements cannot be obtained.
This is because simultaneously modeling asymmetric absorption features and multi-component emission lines
is highly uncertain and often leads to non-physical or unstable fitting outcomes.

\begin{figure*}
\centering
\figurenum{3}
\includegraphics[width=\textwidth]{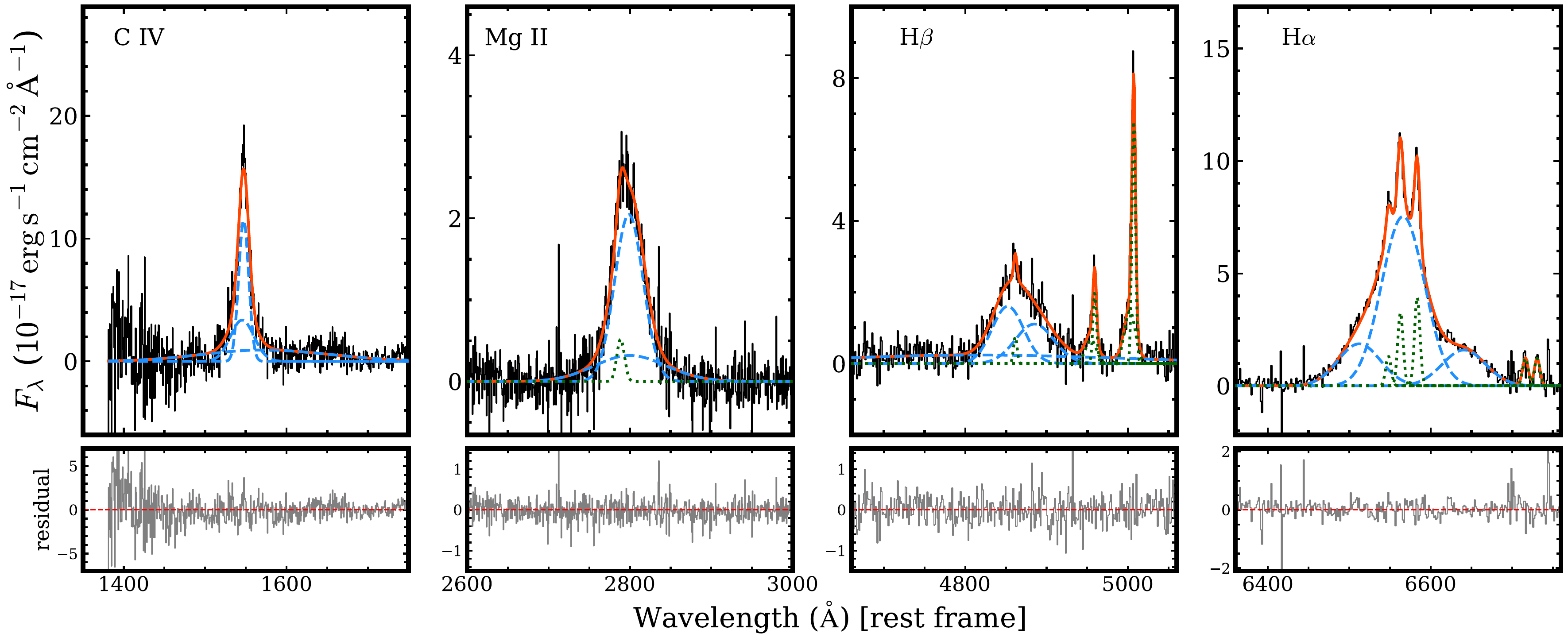}
\caption{
Examples of spectral fitting results for the \ion{C}{4}, \ion{Mg}{2}, H$\beta$, and H$\alpha$ line complexes.
In the upper panels, the black lines show the rest-frame continuum-subtracted spectra.
The dashed blue and dotted green lines represent the broad and narrow emission line models, respectively,
and the solid red lines indicate the combined best-fit model of the broad and narrow lines.
The lower panels show the fitting residuals.
\label{fig:line}}
\end{figure*}

Figure \ref{fig:line} shows examples of the line fitting results. 
The FWHM values of the broad H$\alpha$, H$\beta$, \ion{Mg}{2}, and \ion{C}{4} lines
were corrected for the effects of instrumental spectral resolution.
This correction was applied using the relation of
$\rm FWHM^2 = \left( FWHM_{fit} \right)^2 - \left( FWHM_{inst} \right)^2$,
where $\rm FWHM_{fit}$ is the measured FWHM from the fit,
and $\rm FWHM_{inst}$ is the instrumental resolution.
Note that the instrumental resolution of the DESI spectra corresponds to a spectral resolution of 1.8\,$\rm \AA{}$.
For the spectra from \cite{shim13}, the Hectospec/MMT has a spectral resolution of 6.2\,$\rm \AA{}$,
while the Hydra/WYIN provides a resolution of 5.7\,$\rm \AA{}$.
We then used only reliable FWHM measurements,
selecting those with $\rm FWHM/FWHM_{err} > 3$, 
where $\rm FWHM_{err}$ is the uncertainty from the fit. 

For some objects, multiple broad emission lines were available for FWHM measurements.
Specifically, 205 objects have both \ion{C}{4} and \ion{Mg}{2} FWHM measurements,
47 have both \ion{Mg}{2} and H$\beta$, and 19 have both H$\beta$ and H$\alpha$.
The corresponding rms scatters of the logarithmic FWHM differences are 0.119, 0.100, and 0.105\,dex, respectively,
which are slightly larger than those reported by \cite{bisogni17} (0.070, 0.068, and 0.043\,dex),
but comparable to the $\sim$0.1\,dex scatter between the H$\alpha$ and H$\beta$ line reported by \cite{greene05}.
Nevertheless, our rms scatters are smaller than those obtained for SDSS quasars,
which show rms scatters of 0.211, 0.129, and 0.131\,dex when only objects with S/N$>3$ are considered \citep{rakshit20}.
The larger \ion{C}{4}-\ion{Mg}{2} scatter in \cite{rakshit20} is plausibly due to the inclusion of outflow-affected \ion{C}{4} lines,
whereas such outflow-affected lines were excluded from our line fitting. 

In total, we obtained reliable FWHM measurements for 585 objects:
326 for \ion{C}{4}, 420 for \ion{Mg}{2}, 68 for H$\beta$, and 43 for H$\alpha$.
The redshift coverage for each line is
$z=\mathrm{1.42}$--$\mathrm{4.32}$ for \ion{C}{4},
$\mathrm{0.32}$--$\mathrm{2.39}$ for \ion{Mg}{2},
$\mathrm{0.09}$--$\mathrm{0.98}$ for H$\beta$, and
$\mathrm{0.09}$--$\mathrm{0.46}$ for H$\alpha$.
Overall, the sample spans $0.09 \leq z \leq 4.32$,
and most objects ($\sim 80$\,\%) are located within $0.5 \leq z \leq 2.5$.
These measurements are summarized in Table \ref{tbl:info_est}. 

\section{Result} \label{sec:result}
\subsection{Bolometric luminosity} \label{sec:Lbol}

In this subsection,
we derived the bolometric luminosities of our sample using the $L_{\rm MIR}$-based estimators \citep{kim23}.
For this, we used the extinction-corrected $L_{\rm 3.4}$ and $L_{\rm 4.6}$ values obtained in Section \ref{sec:SED}.
The $L_{\rm MIR}$-based $L_{\rm bol}$ estimators are
\begin{equation}
	\begin{aligned}
	& \log \left( \frac{L_{\rm bol}}{10^{44}\,{\rm erg~s^{-1}}} \right) \\
	& =(0.722 \pm 0.021) + (0.993 \pm 0.031) \log \left( \frac{L_{\rm 3.4}}{10^{44}\,{\rm erg~s^{-1}}} \right) \label{eqn:Lbol_34}
	\end{aligned}
\end{equation}
and
\begin{equation}
	\begin{aligned}
		& \log \left( \frac{L_{\rm bol}}{10^{44}\,{\rm erg~s^{-1}}} \right) \\
		& =(0.739 \pm 0.021) + (0.993 \pm 0.031) \log \left( \frac{L_{\rm 4.6}}{10^{44}\,{\rm erg~s^{-1}}} \right), \label{eqn:Lbol_46}
	\end{aligned}
\end{equation}
with an intrinsic scatter of 0.126\,dex for both relations.
Note that deriving $L_{\rm bol}$ from a full SED covering X-ray to IR wavelengths (e.g., \citealt{krawczyk13}) is the most reliable estimate.
However, this is not feasible for all sources given the wide redshift range and photometric coverage of our sample.
Therefore, we adopted $L_{\rm MIR}$ derived from the SED fitting \citep{kim23} as a proxy for $L_{\rm bol}$.

Using these  $L_{\rm MIR}$-based estimators,
we obtained $L_{\rm bol}$ values based on $L_{\rm 3.4}$ and $L_{\rm 4.6}$ for 609 and 532 objects, respectively.
Since both $L_{\rm 3.4}$ and $L_{\rm 4.6}$ were derived from the single AGN component obtained through a common SED fit,
the resulting $L_{\rm bol}$ values are effectively the same.
Any difference is negligible in practice compared to the intrinsic scatter of $\sim$0.13\,dex.
Using these measurements, we obtained $L_{\rm bol}$ values for 639 objects,
each with at least one available measurement of $L_{\rm 3.4}$ or $L_{\rm 4.6}$.
These $L_{\rm bol}$ values are listed in Table \ref{tbl:prop}
and span a wide range from $10^{43.20}$ to $10^{47.27}\,{\rm erg~s^{-1}}$,
with mean and median values of $10^{45.49}$ and $10^{45.55}\,{\rm erg~s^{-1}}$, respectively,
and a standard deviation of 0.61\,dex.
Note that the uncertainties in the $L_{\rm bol}$ values were estimated by
taking into account both the uncertainties from the $L_{\rm MIR}$ measurements
and those in the parameters of the $L_{\rm bol}$ estimators.

\begin{deluxetable}{cccc}
	\tabletypesize{\scriptsize}
	\tablecolumns{4}
	\tablewidth{0pt}
	\tablenum{2}
	\tablecaption{BH Properties of Type-1 AGNs in the NEP-Wide Field\label{tbl:prop}}
	\tablehead{
		\colhead{$\it AKARI$ ID}&
		\colhead{$\log \left( \frac{L_{\rm bol}}{\rm erg~s^{-1}} \right)$\tablenotemark{a}}& 
		\colhead{$\log \left( \frac{M_{\rm BH}}{M_{\odot}} \right)$\tablenotemark{a}}&
		\colhead{$\log \left( \lambda_{\rm Edd} \right)$}
	}
	\startdata
	133109  &  $44.55 \pm 0.01$  &  $7.97 \pm 0.04$  &  $-1.52 \pm 0.04$  \\ 
	135036  &  $45.43 \pm 0.04$  &  $8.48 \pm 0.21$  &  $-1.16 \pm 0.17$  \\ 
	44872  &  $43.71 \pm 0.01$  &  $7.97 \pm 0.08$  &  $-2.36 \pm 0.08$  \\ 
	134539  &  --  &  --  &  --  \\
	133581  &  $44.53 \pm 0.01$  &  $7.81 \pm 0.04$  &  $-1.38 \pm 0.04$  \\ 
	46186  &  $44.90 \pm 0.03$  &  --  &  --  \\
	133141  &  $44.82 \pm 0.02$  &  $8.15 \pm 0.05$  &  $-1.44 \pm 0.05$  \\ 
	132335  &  --  &  --  &  --  \\ 
	132310  &  $45.73 \pm 0.05$  &  $8.51 \pm 0.04$  &  $-0.88 \pm 0.06$  \\ 
	132768  &  $43.59 \pm 0.01$  &  --  &  --  \\
	133094  &  --  &  --  &  --  \\ 
	133299  &  $46.28 \pm 0.07$  &  --  &  --  \\
	132336  &  $44.02 \pm 0.01$  &  --  &  --  \\
	53113  &  --  &  --  &  --  \\ 
	132128  &  --  &  --  &  --  \\
	132658  &  --  &  --  &  --  \\  
	132883  &  $44.31 \pm 0.01$  &  --  &  --  \\ 
	\enddata
	\tablenotetext{a}{Representative $L_{\rm bol}$ and $M_{\rm BH}$ values (see Sections \ref{sec:Lbol} and \ref{sec:MBH}).
	\tablenotetext{a}{The listed values do not include the intrinsic rms scatters of the $L_{\rm bol}$ and $M_{\rm BH}$ estimators,
	which are $\sim$0.13\,dex and $\sim$0.23\,dex, respectively.}
	}
	\tablecomments{Column (1): $\it AKARI$ ID. Column (2): Logarithmic bolometric luminosity in units of $\rm erg~s^{-1}$.
		Column (3): Logarithmic BH mass in units of $M_{\odot}$.
		Column (4): Logarithmic Eddington ratio.
		This table is fully available in a machine-readable format.}
\end{deluxetable}

\subsection{Black Hole Mass} \label{sec:MBH}

In this subsection, we estimate the BH masses using the $L_{\rm MIR}$-based estimators.
For objects with available FWHM measurements of the H$\beta$ or H$\alpha$ lines,
we applied the $L_{\rm MIR}$-based $M_{\rm BH}$ estimators presented in \cite{kim23}.
These $L_{\rm MIR}$-based $M_{\rm BH}$ estimators can be expressed as
\begin{equation}
	\begin{aligned}
		\log \left( \frac{M_{\rm BH}}{M_{\rm \odot}} \right) = & \alpha + \beta \log \left( \frac{L_{\rm Cont}}{\rm 10^{44}\,erg~s^{-1}} \right)\\
		&+\gamma \log \left( \frac{\rm FWHM_{Line}}{\rm 10^{3}\,km~s^{-1}} \right), \label{eqn:MBH_Bal}
	\end{aligned}
\end{equation}
where the coefficients $\alpha$, $\beta$, and $\gamma$ are listed in Table \ref{tbl:MBH_est}.

\begin{deluxetable}{cccc}
	\tabletypesize{\scriptsize}
	\tablecolumns{4}
	\tablewidth{0pt}
	\tablenum{3}
	\tablecaption{Coefficients of $L_{\rm MIR}$-based $M_{\rm BH}$ estimators\label{tbl:MBH_est}}
	\tablehead{
		\colhead{No.}&	\colhead{$L_{\rm Cont}$}&	\colhead{$\rm FWHM_{Line}$}& \colhead{$\alpha$}
	}
	\startdata
	A&	$L_{\rm 3.4}$&	\ion{C}{4}&		 6.503$\pm$0.172\\
	B&	$L_{\rm 3.4}$&	\ion{Mg}{2}&	6.939$\pm$0.116\\
	C&	$L_{\rm 3.4}$&	H$\beta$&		6.906$\pm$0.023\\
	D&	$L_{\rm 3.4}$&	H$\alpha$&	   6.964$\pm$0.060\\
	\hline
	E&	$L_{\rm 4.6}$&	\ion{C}{4}&	      6.512$\pm$0.172\\
	F&	$L_{\rm 4.6}$&	\ion{Mg}{2}&	 6.948$\pm$0.116\\
	G&	$L_{\rm 4.6}$&	H$\beta$&       6.914$\pm$0.023\\
	H&	$L_{\rm 4.6}$&	H$\alpha$&	    6.973$\pm$0.060\\
	\enddata
	\tablecomments{For all lines, $\beta$ is 0.478$\pm$0.016,
	while $\gamma$ is 2 for H$\beta$ and 2.06$\pm$0.06 for the others.}
\end{deluxetable}

For objects with FWHM measurements of the \ion{Mg}{2} and/or \ion{C}{4} lines,
we modified the H$\alpha$ line-based $M_{\rm BH}$ estimators (i.e., D and H in Table \ref{tbl:MBH_est})
by applying the empirical relations between $\rm FWHM_{H\alpha}$ and 
FWHMs of the \ion{Mg}{2} and/or \ion{C}{4} lines from \cite{bisogni17}.
\cite{bisogni17} obtained optical to NIR spectra simultaneously for six quasars at $z$$\sim$2.2
using the X-shooter spectrograph on the Very Large Telescope,
and derived the following relations:
\begin{equation}
	\begin{aligned}
		\log & \left( \frac{\rm FWHM_{H\alpha}}{\rm 1000\,km~s^{-1}} \right)\\
		&= \log \left( \frac{\mathrm{FWHM}_{\mathrm{Mg\,\textsc{ii}}}}{\rm 1000\,km~s^{-1}} \right) - (0.012 \pm 0.048) \label{eqn:FWHM_Ha-MgII}
	\end{aligned}
\end{equation}
and
\begin{equation}
	\begin{aligned}
		\log & \left( \frac{\rm FWHM_{H\alpha}}{\rm 1000\,km~s^{-1}} \right)\\
		&= \log \left( \frac{\mathrm{FWHM}_{\mathrm{C\,\textsc{iv}}}}{\rm 1000\,km~s^{-1}} \right) - (0.224 \pm 0.078). \label{eqn:FWHM_Ha-CIV}
	\end{aligned}
\end{equation}
The obtained $M_{\rm BH}$ estimators based on the \ion{Mg}{2} line are summarized as B and F in Table \ref{tbl:MBH_est},
while those based on the \ion{C}{4} line are listed as A and E.

Note that the H$\alpha$-based $M_{\rm BH}$ estimator \citep{kim23} has an rms scatter of 0.204\,dex.
Considering the $\sim$0.11\,dex scatter in FWHM measurements between different emission lines as described in Section \ref{sec:line},
the intrinsic scatters of the BH mass estimators in Table \ref{tbl:MBH_est} become $\sim$0.23\,dex.

Using these $M_{\rm BH}$ estimators, we derived BH masses for 212, 352, 64, and 37 objects
using the \ion{C}{4}, \ion{Mg}{2}, H$\beta$, and H$\alpha$ lines, respectively.
As with the representative $L_{\rm bol}$ values described in Section \ref{sec:Lbol},
we constructed representative $M_{\rm BH}$ values for each object by combining the available measurements.
Since the $L_{\rm 3.4}$ and $L_{\rm 4.6}$ values were derived from a common SED fit,
the resulting $M_{\rm BH}$ estimates are identical and were treated as a single measurement.
Furthermore, when multiple line-based estimates are available,
we prioritized those derived from the H$\beta$ line, followed by the \ion{Mg}{2}, H$\alpha$, and \ion{C}{4} lines.
The representative BH masses, listed in Table \ref{tbl:prop}, were obtained for 447 objects and
span a wide range from $10^{7.29}$ to $10^{9.67}\,M_{\odot}$.
The mean and median $M_{\rm BH}$ values are $10^{8.65}$ and $10^{8.67}\,M_{\odot}$, respectively,
and the standard deviation is 0.42\,dex.
Note that the uncertainties in the $M_{\rm BH}$ values were estimated
by taking into account the uncertainties of the $L_{\rm MIR}$ and FWHM measurements,
together with those in the parameters of the $M_{\rm BH}$ estimators.

\subsection{Eddington Ratio} \label{sec:REdd}

In this subsection, we derived the Eddington ratios (hereafter, $\lambda_{\rm Edd}$)
defined as $\lambda_{\rm Edd} \equiv L_{\rm bol}$/$L_{\rm Edd}$, where $L_{\rm Edd}$ is the Eddington luminosity.
We used the representative values of $L_{\rm bol}$ and $M_{\rm BH}$
to compute the $\lambda_{\rm Edd}$ values, resulting in measurements for 447 objects.
The measured $\lambda_{\rm Edd}$ values are summarized in Table \ref{tbl:prop}.

\begin{figure}
\centering
\figurenum{4}
\includegraphics[width=\columnwidth]{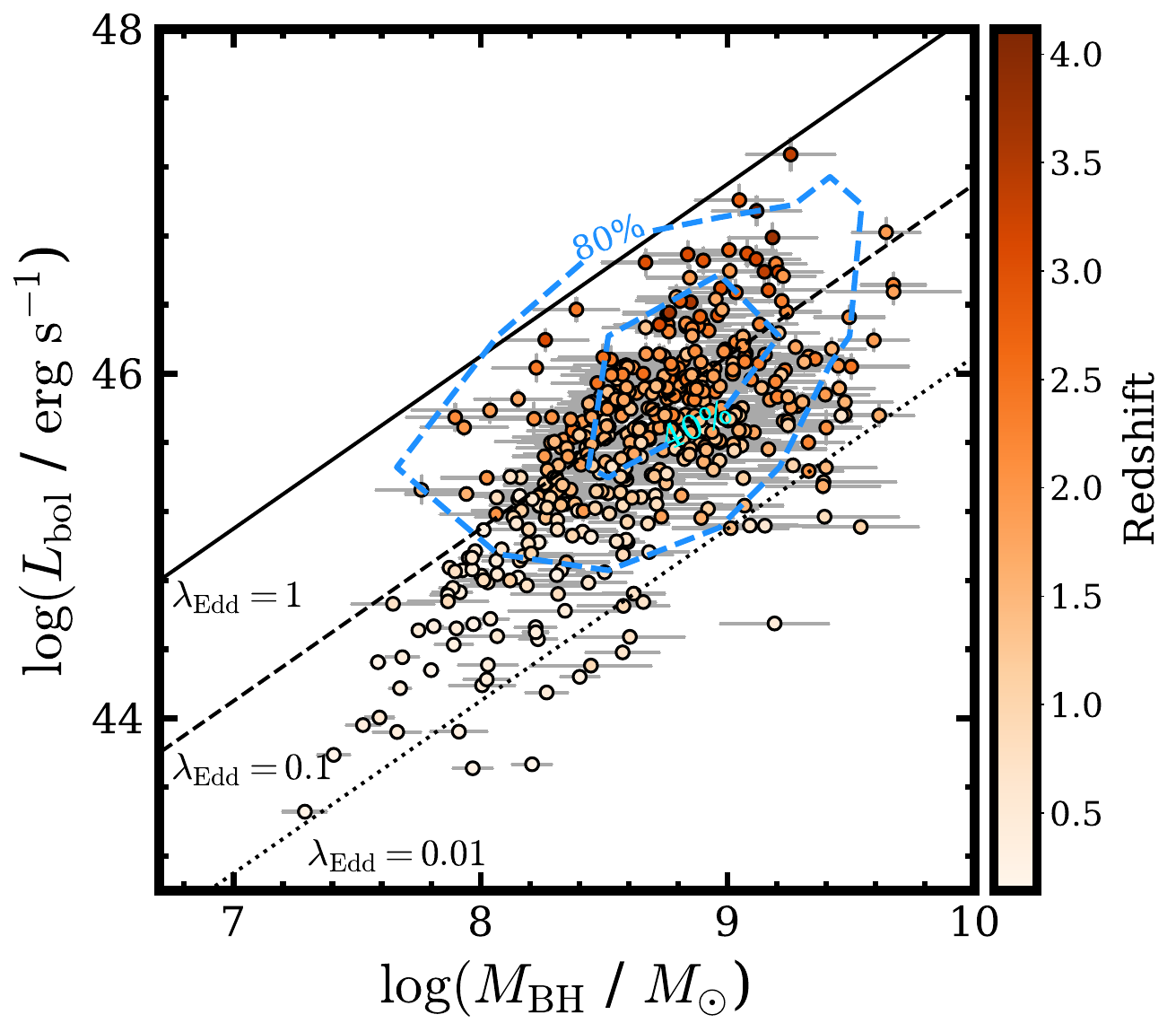}
\caption{
Distribution of BH masses and bolometric luminosities.
Circles represent the representative BH masses and bolometric luminosities of Type-1 AGNs in the NEP-Wide field;
their colors indicate redshift. Blue dashed contours indicate the distribution of SDSS DR14 quasars.
The solid, dashed, and dotted lines correspond to $\lambda_{\rm Edd}$ values of 1.0, 0.1, and 0.01, respectively.
\label{fig:MBH-Lbol}}
\end{figure}

Figure \ref{fig:MBH-Lbol} presents the distributions of $L_{\rm bol}$ and $M_{\rm BH}$,
with reference lines corresponding to $\lambda_{\rm Edd} =1$, 0.1, and 0.01.
All of the derived $\lambda_{\rm Edd}$ values are lower than 1,
with $\log(\lambda_{\rm Edd})$ ranging from -2.74 to -0.08.
The mean and median $\log(\lambda_{\rm Edd})$ values are -1.17 and -1.12, respectively,
and the standard deviation is 0.43\,dex.
Moreover, the $\lambda_{\rm Edd}$ distributions of our sample
is broadly consistent with that of the SDSS DR14 quasars, as shown in Figure \ref{fig:MBH-Lbol}.

\section{Discussion} \label{sec:discussion}
\subsection{Prevalence of Dust-obscured AGNs} \label{sec:DOAGN_Frac}

The fraction of dust-obscured AGNs provides key insights into the cosmic evolution of both dust-obscured and unobscured AGN populations.
Moreover, a substantial dust-obscured fraction
(i) suggests the existence of a significant population of AGNs that have been missed in optical AGN surveys,
and (ii) underscores the critical importance of correcting dust extinction 
when investigating the nature of AGNs (e.g., \citealt{kim23,kim25}).
Additionally, the dust-obscured fraction may provide indirect evidence for a connection between 
galaxy mergers and galaxy evolution (e.g., \citealt{hopkins06,kocevski15,kim20,lamarca24}).
Some X-ray-based studies have found the dust-obscured fraction to be $\geq 50$\,\% (e.g., \citealt{martinez05}),
though this fraction can vary depending on the redshift and AGN luminosity (e.g., \citealt{akylas06,polletta08,peca23}).
In addition, even optical-based AGN surveys also have non-negligible dust-obscured fractions.
For instance, \cite{kim23} reported that a significant fraction ($\sim$15\,\%) of SDSS quasars are dust-obscured,
and that their $L_{\rm bol}$ values derived from the optical continuum luminosities can be underestimated by $\sim$50\,\%.

In this subsection, we investigated the dust-obscured AGN fraction for the Type-1 AGNs in the NEP-Wide field.
For this analysis, we used a sample of 639 objects with the $E(B-V)$ values obtained from the SED fitting, as described in Section \ref{sec:SED}.
Based on these $E(B-V)$ values,
we classified objects with $E(B-V) > 0.1$ as dust-obscured AGNs, and those with $E(B-V) \leq 0.1$ as unobscured AGNs.
Note that we tried to test our dust-obscured AGN classification using the Balmer decrement.
However, only three dust-obscured AGNs have both H$\beta$ and H$\alpha$ flux measurements.
Moreover, two of them have $E(B-V) \lesssim 0.2$; given that the rms scatter between $E(B-V)$
derived from the Balmer decrement and that from SED fitting is $\sim$0.2,
we could not robustly validate our SED-based classification due to the small sample size.

\begin{figure}
	\centering
	\figurenum{5}
	\includegraphics[width=\columnwidth]{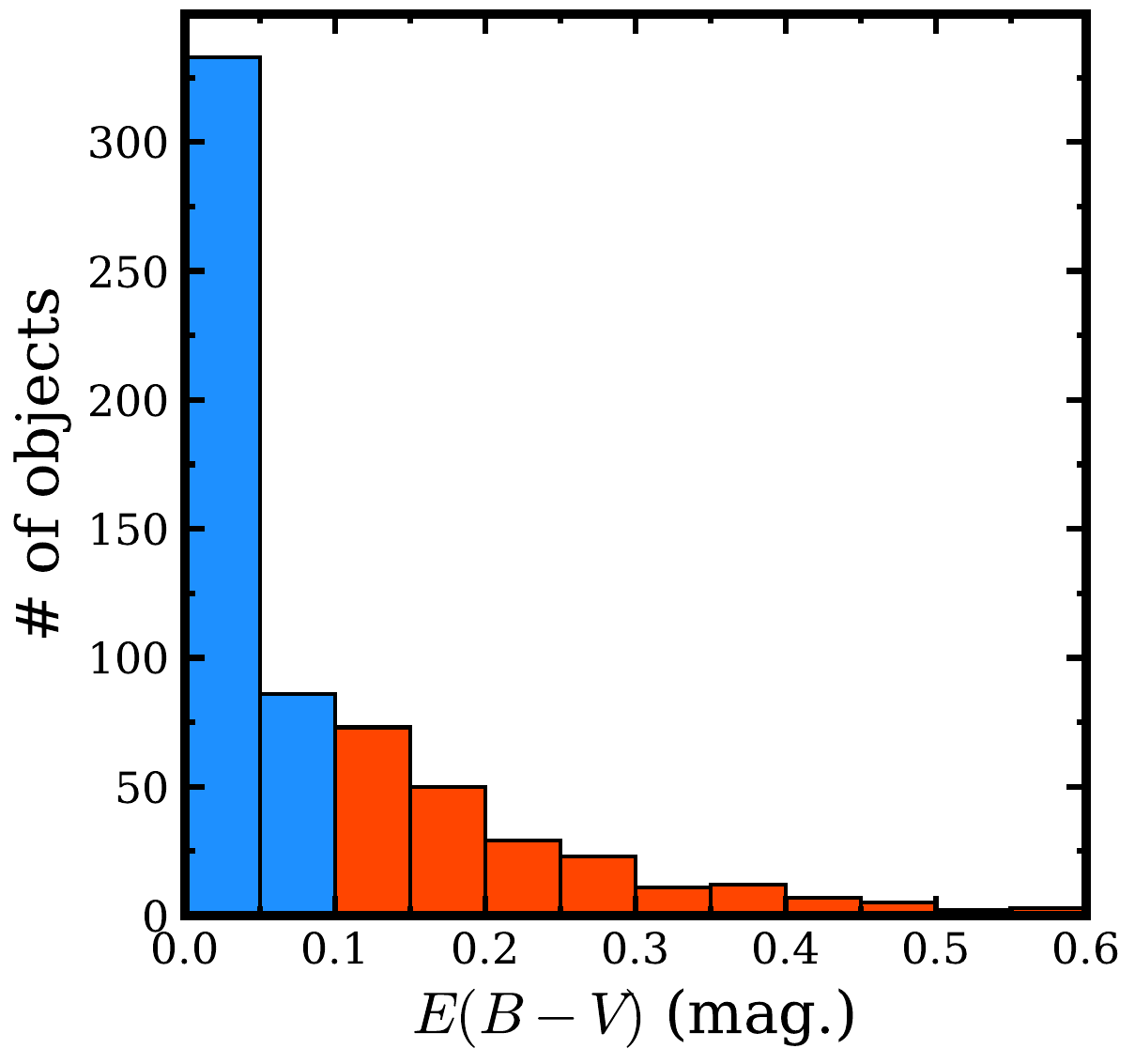}
	\caption{
		Distribution of $E(B-V)$ values derived from SED fitting.
		The blue and red histograms represent unobscured [$E(B-V) \leq 0.1$] and dust-obscured [$E(B-V) > 0.1$] AGNs, respectively.
		\label{fig:EBV_dist}}
\end{figure}

\begin{figure*}
\centering
\figurenum{6}
\includegraphics[width=\textwidth]{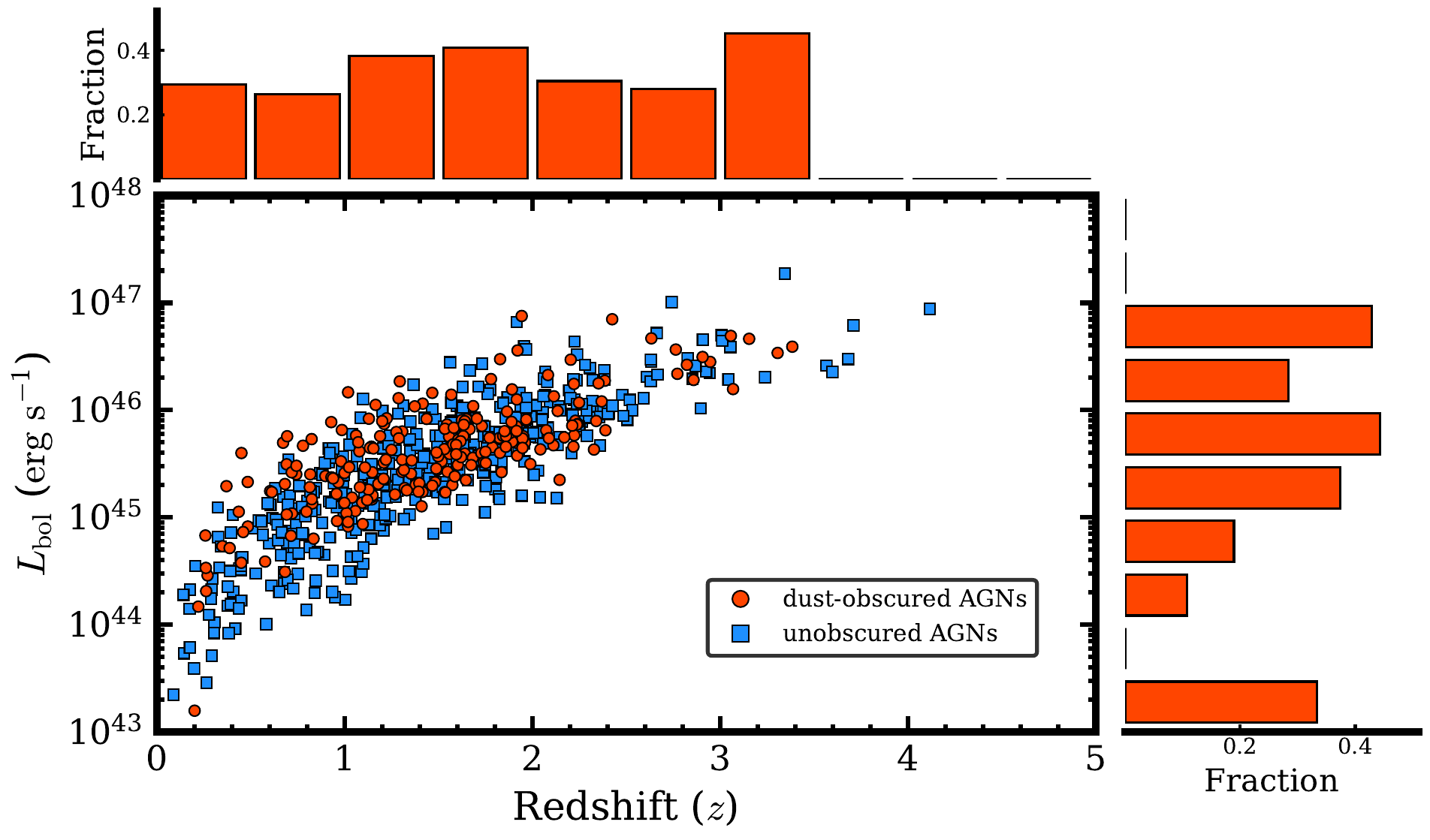}
\caption{
Distributions of bolometric luminosity versus redshift for dust-obscured and unobscured AGNs.
Red circles and blue squares represent dust-obscured and unobscured AGNs, respectively.
The top and right panels show the redshift and bolometric luminosity distributions of the dust-obscured fraction, respectively.
\label{fig:RQ_Sample}}
\end{figure*}

Figure \ref{fig:EBV_dist} shows the distribution of the $E(B-V)$ values.
As a result, 220 and 419 objects were classified as dust-obscured and unobscured AGNs, respectively,
and the redshifts of the dust-obscured AGNs span a wide range from 0.20 to 3.38, as shown in Figure \ref{fig:RQ_Sample}.
Note that this $E(B-V)$ criterion has been widely adopted in previous studies (e.g., \citealt{glikman07,kim15a,kim23,kim24b}).
However, due to the significant uncertainties in $E(B-V)$ measurements,
some studies (e.g., \citealt{kim18b}) adopted higher thresholds to classify dust-obscured AGNs.

Using these dust-obscured and unobscured AGNs, we estimated the dust-obscured fraction,
defined as the number of dust-obscured AGNs divided by the total number of both dust-obscured and unobscured AGNs,
and found a dust-obscured fraction of 34\,\%.
However, the dust-obscured fraction can vary depending on how dust-obscured AGNs are defined.
\cite{kim23} defined dust-obscured AGNs as those whose extinction-corrected $L_{\rm bol}$ is more than
1.5 times the $L_{\rm bol}$ derived from the uncorrected L5100.
Using this definition, objects with $E(B-V) \gtrsim 0.13$ are classified as dust-obscured AGNs,
and the corresponding dust-obscured fraction decreases to $\sim$28\,\%.
This fraction is still higher than that of the SDSS quasars \citep{kim23}, which is 16\,\%.
While this discrepancy can result from various factors such as photometric depth and the definition of the dust-obscured fraction,
(i) the optical-to-IR color-based sample selection adopted in the DESI and
(ii) the tendency of $\it{AKARI}$-detected AGNs to favor a more obscured population are suspected to be the primary factors.
Furthermore, \cite{kim18b} reported that the typical rms scatter in the $E(B-V)$ measurements is $\sim$0.2,
which could have introduced a minor bias into the derived dust-obscured fraction.

Although the significant dust-obscured fraction was found among the Type-1 AGNs detected in the optical and MIR surveys,
this estimate may still suffer from various selection biases, particularly as heavily obscured AGNs tend to be missed.  
To mitigate such biases, we further estimated the dust-obscured fraction using a subsample of X-ray detected objects from our sample. 
The NEP-Deep field, located within the NEP-Wide field, was observed by a 300\,ks \textit{Chandra} survey \citep{krumpe15}.
This survey detected 457 X-ray sources, of which 33 were matched to our sample with available $E(B-V)$ measurements.
Based on the X-ray detected AGNs, we derived a dust-obscured fraction of 27\,\%.
Furthermore, considering the AGNs detected in the hard X-ray (2--7\,keV) band,
we obtained a comparable fraction of 30\,\%.
These estimates are broadly consistent with the dust-obscured fraction obtained from the optical-MIR detected AGNs.

To further examine the dust-obscured fraction as a function of redshift,
we selected AGNs with bolometric luminosities in the range $10^{44.5}$--$10^{46.5}\,{\rm erg~s^{-1}}$,
yielding a sample of 569 objects (206 dust-obscured and 363 unobscured AGNs),
which is sufficiently large to provide statistically robust results.
To alleviate luminosity-dependent biases, we further divided this sample into four $L_{\rm bol}$ bins:
$10^{44.5}$--$10^{45.0}\,{\rm erg~s^{-1}}$, $10^{45.0}$--$10^{45.5}\,{\rm erg~s^{-1}}$,
$10^{45.5}$--$10^{46.0}\,{\rm erg~s^{-1}}$, and $10^{46.0}$--$10^{46.5}\,{\rm erg~s^{-1}}$.
Within these $L_{\rm bol}$ bins, 14, 67, 98, and 27 dust-obscured AGNs were selected,
while 60, 112, 123, and 68 unobscured AGNs were chosen, respectively.

\begin{figure}
\centering
\figurenum{7}
\includegraphics[width=\columnwidth]{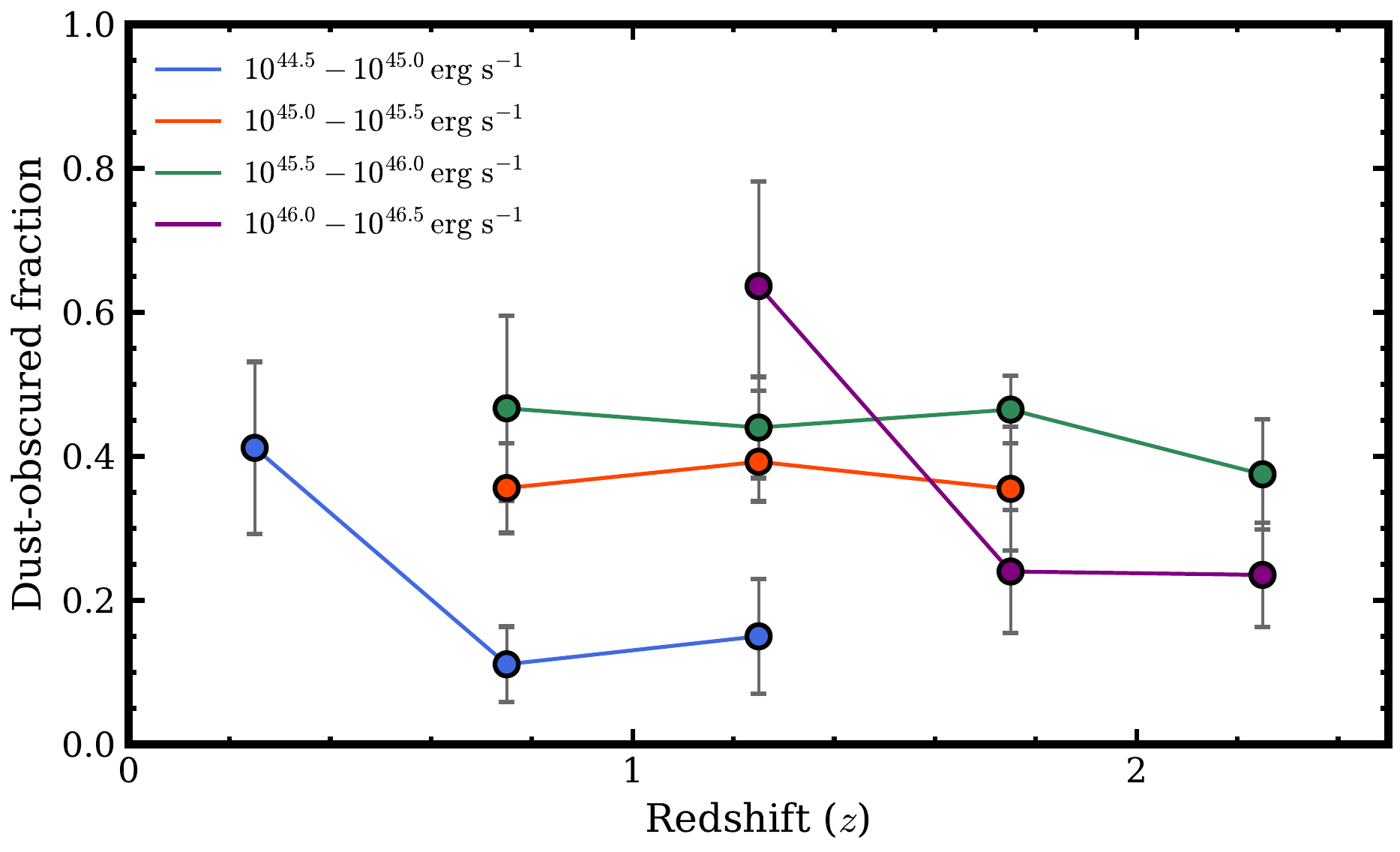}
\caption{
Dust-obscured AGN fraction as a function of redshift in four bolometric luminosity subsamples.
The blue, red, green, and purple circles and lines denote the fractions for $L_{\rm bol}$ ranges of
$10^{44.5}$--$10^{45.0}$, $10^{45.0}$--$10^{45.5}$, $10^{45.5}$--$10^{46.0}$, and $10^{45.5}$--$10^{46.0}$, respectively.
Only cases where the number of AGNs is at least ten are shown to avoid the effects of small number statistics.
\label{fig:RQ_z}}
\end{figure}

For these subsamples, the derived dust-obscured fractions are comparable,
as shown in Figure \ref{fig:RQ_z}.
Since there are only a few AGNs in some redshift bins,
bins fewer than 10 objects are not presented in Figure \ref{fig:RQ_z}.
The dust-obscured fraction is typically $\sim$40\,\% (ranging from $\sim$10\,\% to $\sim$65\,\%) in the redshift range $0.5 < z < 2.0$,
which is consistent with several previous studies (e.g., \citealt{akylas06}).
The subsamples with $L_{\rm bol}$ between $10^{45.0}$--$10^{45.5}\,{\rm erg~s^{-1}}$ and $10^{45.5}$--$10^{46.0}\,{\rm erg~s^{-1}}$
exhibit no clear redshift-dependent trend in the dust-obscured fraction.
In contrast, the subsamples with $L_{\rm bol}$ in the ranges $10^{44.5}$--$10^{45.0}\,{\rm erg~s^{-1}}$ and $10^{46.0}$--$10^{46.5}\,{\rm erg~s^{-1}}$
show a declining trend beyond $z \gtrsim 0.5$ and $z \gtrsim 1.5$, respectively.
Such a trend needs to be verified with a larger sample, 
given the limited number of dust-obscured AGNs in these bins ($<$10 objects in each bin)
and the potential effects of selection biases.
Note that even when using subsamples with narrower $L_{\rm bol}$ or redshift bins,
no significant new trends were revealed.

Note that, since the DESI quasars were selected based on the optical-to-IR colors,
the sample cannot be considered complete for dust-obscured objects.
Consequently, the redshift trends could be biased.
As spectroscopic surveys of the NEP-Wide field become more complete through SPHEREx and other upcoming missions,
both the number and completeness of identified dust-obscured AGNs will increase, 
leading to more statistically robust and comprehensive results for the dust-obscured fraction and its redshift dependence. 

\subsection{Spectral Energy Distribution of Dust-obscured AGNs} \label{sec:SED_shape}

We compared the SED shapes of the dust-obscured and unobscured AGNs.
The SEDs for all AGNs were constructed from the multi-band photometry used in Section \ref{sec:SED}.
Owing to the closely spaced wavelengths of the $\it{AKARI}$, WISE, and $\it{Spitzer}$ IR photometric data points,
such photometry provides an ideal data set for SED shape analysis in the IR region.
Using the $\it{K}$-band wavelength, where the effect of dust extinction is expected to be minimal, as the reference point,
we normalized the SEDs and then median-combined them within each type for comparison.
The 16th and 84th percentiles of the combined SEDs were used as the $1\sigma$ confidence level.

\begin{figure}
\centering
\figurenum{8}
\includegraphics[width=\columnwidth]{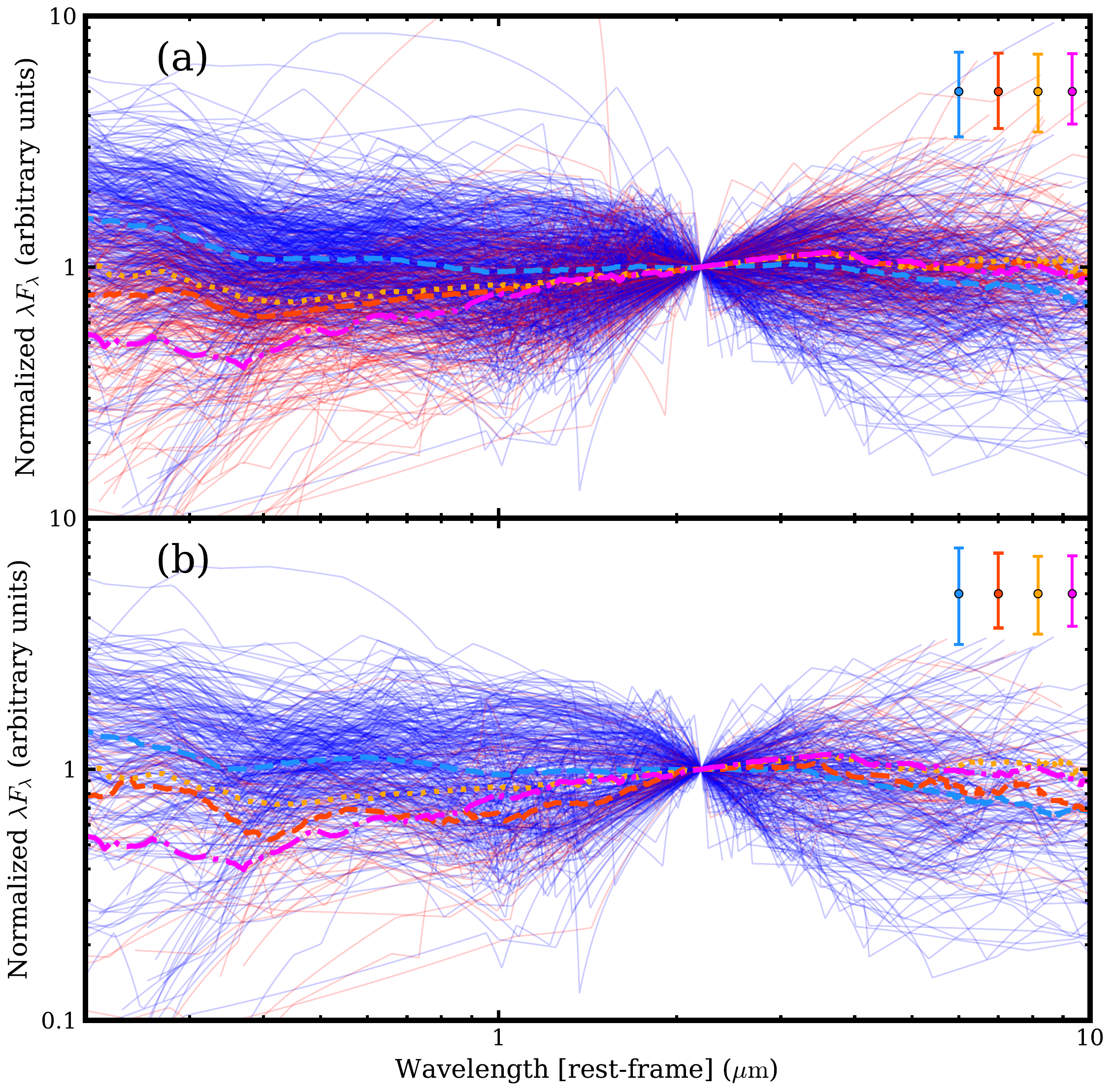}
\caption{
Normalized SED shapes of dust-obscured and unobscured AGNs.
Panels (a) and (b) show the entire AGN sample and
those of AGNs with $L_{\rm bol}$ in the range of $10^{45}$--$10^{46}\,{\rm erg~s^{-1}}$, respectively.
Red and blue solid lines represent individual dust-obscured and unobscured AGN SEDs, respectively,
normalized at the $\it{K}$-band wavelength.
The red and blue dashed lines indicate the median-combined SEDs for the two populations. 
The orange dotted and magenta dot-dashed lines represent the median-combined SEDs of
moderately obscured ($0.1 < E(B-V) < 0.2$) and heavily obscured ($E(B-V) > 0.2$) AGNs, respectively.
The red and blue error bars in the top-right corner show
the representative 1$\sigma$ confidence levels of the combined SED shapes
for dust-obscured and unobscured AGNs, respectively.
The orange and magenta error bars correspond to the same confidence levels for
the moderately and heavily obscured AGN subsamples, respectively.
\label{fig:RQ_SED}}
\end{figure}

The comparison of the SED shapes for the dust-obscured and unobscured AGNs is presented in Figure \ref{fig:RQ_SED}.
In the optical region, they differ significantly:
at 5100\,$\rm \AA{}$, the combined SED of the unobscured AGNs is 57\,\% higher than that of the dust-obscured AGNs.
Considering the $1\sigma$ confidence level for the combined SED of the dust-obscured AGNs (37\,\%),
this difference between the two types is statistically significant.
Furthermore, by dividing the dust-obscured AGNs into moderately obscured ($0.1 < E(B-V) \leq 0.2$)
and heavily obscured ($E(B-V) > 0.2$) AGNs, we find that
the SED differences relative to the unobscured AGNs become more pronounced as $E(B-V)$ increases.
In contrast, in the IR region at wavelengths of 3.4\,$\mu$m and 4.6\,$\mu$m,
the combined SED of the unobscured AGNs is 91\,\% and 92\,\% of that of the dust-obscured AGNs,
differences that are not statistically significant given the corresponding $1\sigma$ uncertainties of 18\,\% and 25\,\%, respectively.
Moreover, the overall IR SED shapes of the two AGN populations show no substantial differences,
nor do they exhibit any noticeable change with increasing $E(B-V)$.

Since (i) the IR SED shapes can be affected by dust torus properties (e.g., \citealt{kim15b,son23})
and (ii) the geometry of the dust torus is known to be related to luminosity (e.g., \citealt{toba13,mkim24}),
we restricted our sample to AGNs with $L_{\rm bol}$ in the range of $10^{45}$--$10^{46}\,{\rm erg~s^{-1}}$
and compared their SED shapes.
Figure \ref{fig:RQ_SED} presents the SED shapes for the $L_{\rm bol}$-limited dust-obscured and unobscured AGNs,
and the overall trends are broadly consistent with those obtained from the entire sample. 
In addition, a similar analysis performed for the moderately obscured and heavily obscured AGN subsamples
yielded results in agreement with the same trend.

The optical SED shapes of the dust-obscured AGNs are thought to be affected by dust extinction,
whereas the similarity of their IR SED shapes to those of the unobscured AGNs 
suggests that dust torus properties such as the covering factor and the inclination angle 
are not significantly different between the two AGN populations.
This similarity further provides strong support
for the applicability of the $L_{\rm MIR}$-based $L_{\rm bol}$ and $M_{\rm BH}$ estimators \citep{kim23}, 
established with unobscured AGNs, to dust-obscured AGNs.

\subsection{Dust Extinction Correction: A Key Element in AGN Studies} \label{sec:Ext_Corr}

Consistent with previous X-ray-based studies (e.g., \citealt{martinez05}),
we found that the Type-1 AGNs in the NEP-Wide field exhibit a substantial ($\sim$35\,\%) dust-obscured fraction.
Moreover, even SDSS quasars, which are typically regarded as relatively unaffected by dust extinction,
show a non-negligible dust-obscured fraction of $\sim$15\,\% \citep{kim23}.
These results suggest that dust extinction can significantly affect a substantial fraction of AGNs,
including those selected from optical AGN surveys that are generally considered unobscured.
Therefore, as emphasized by \cite{kim25}, 
correcting for dust extinction should be considered as a fundamental step in AGN studies, even for optically selected sources. 

In this subsection, we investigated how the use of UV or optical continuum luminosities without extinction correction
can lead to an underestimation of bolometric luminosities.
For this purpose, we derived $L_{\rm bol}$ values using 
UV/optical continuum luminosity-based estimators without applying dust extinction correction, 
and compared them with our $L_{\rm MIR}$-based $L_{\rm bol}$ values.
To derive the $L_{\rm bol}$ values from UV/optical continuum luminosities,
we measured the monochromatic continuum luminosities at 1450\,$\rm \AA{}$, 3000\,$\rm \AA{}$, and 5100\,$\rm \AA{}$
(hereafter, L1450, L3000, and L5100, respectively),
which have been widely used as proxies for $L_{\rm bol}$ (e.g., \citealt{runnoe12,chen25}).
These UV/optical luminosities were measured using the \texttt{PyQSOFit} code
applied to the DESI DR1 and \cite{shim13} spectra, after correcting for Galactic dust extinction.

Using these UV/optical continuum luminosities, we derived $L_{\rm bol}$ values 
based on the empirical linear relations from \cite{chen25}.
For a direct comparison with the $L_{\rm bol}$ values obtained using the $L_{\rm MIR}$-based estimators from \cite{kim23},
who defined $L_{\rm bol}$ as the integrated luminosity from 10\,KeV to 1\,$\mu$m,
we adopted the linear relations derived for $L_{\rm bol}$ over the same wavelength range.

Note that L1450--$L_{\rm bol}$ relation is sensitive to variations in SED shapes;
consequently, various L1450-based $L_{\rm bol}$ estimators (e.g., \citealt{shen11,runnoe12,chen25}) can yield noticeably different values.
This sensitivity to SED shapes may arise from object-to-object differences in the UV bump,
whose peak position can vary significantly among AGNs
depending on the accretion disk temperature, inclination, and BH mass (e.g., \citealt{malkan83,czerny87,scott04}). 
Hence, for SDSS quasars, the fractional residual, defined as $ \left| \frac{ L_{\rm bol(model)} - L_{\rm bol(true)} }{L_{\rm bol(true)}} \right|$,
has been found to be 34 times larger for L1450 than for L3000, and 2.4 times larger than for L5100 \citep{chen25}.
Therefore, L1450-based $L_{\rm bol}$ estimators yield different results; for instance,
if we adopted the L1450-based estimator from \cite{runnoe12},
the derived $L_{\rm bol}$ values would increase by a factor of $\sim$1.71.

Moreover, to clearly examine the role of dust extinction correction in the comparison,
we restricted the effects from host galaxy contamination.
Previous studies (e.g., \citealt{shen11,kim23}) showed that 
host galaxy contributions are relatively minor (e.g., $\lesssim 20$\,\%) in the UV and MIR, compared to the optical and NIR.
However, the L5100, as an optical continuum luminosity, can be significantly affected by host galaxy contamination,
particularly in low-luminosity AGNs \citep{shen11,jalan23}.
Therefore, in our L5100-based analysis, we included only objects with $\rm L5100 > 10^{44.5}\,erg~s^{-1}$,
since the host galaxy contributions at this luminosity are expected to be limited ($\lesssim$25\,\%; \citealt{shen11,jalan23,ren24}).
Note that, under such conditions, only 20 objects remain (6 dust-obscured and 14 unobscured AGNs),
which is insufficient to obtain statistically robust results.

\begin{figure*}
	\centering
	\figurenum{9}
	\includegraphics[width=\textwidth]{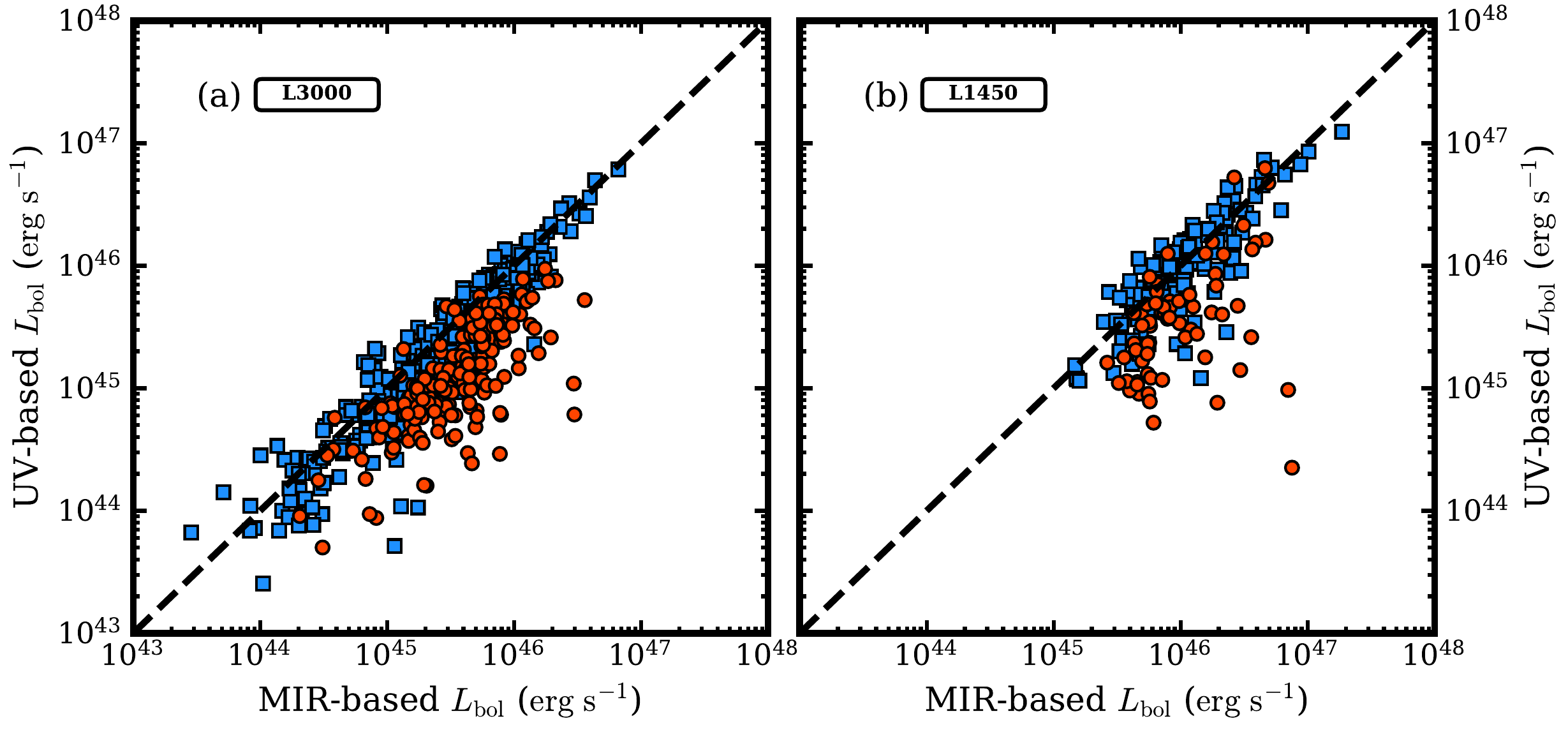}
	\caption{
		Comparison of bolometric luminosities from UV continuum luminosities and those from $L_{\rm MIR}$ values.
		Panels (a) and (b) compare the $L_{\rm MIR}$-based $L_{\rm bol}$ values to those from L3000 and L1450, respectively.
		Red circles and blue squares represent dust-obscured and unobscured AGNs, respectively,
		and the one-to-one relations are shown as the dashed black lines.
		\label{fig:RQ_Lbol_Comp}}
\end{figure*}

We compared the $L_{\rm bol}$ values derived from the $L_{\rm MIR}$ and the UV/optical continuum luminosities
after dividing the sample into dust-obscured and unobscured AGNs, as shown in Figure \ref{fig:RQ_Lbol_Comp}.
However, note that the comparison with the L5100-based $L_{\rm bol}$ was omitted from Figure \ref{fig:RQ_Lbol_Comp}
due to the limited number of objects.
For the unobscured AGNs, the $L_{\rm bol}$ values estimated from UV/optical luminosities
are broadly consistent with those derived from $L_{\rm MIR}$, with $\log [L_{\rm bol}({\rm UV,\,opt})$/$L_{\rm bol}$(${\rm MIR}$)]=
$-0.04 \pm 0.02$, $-0.10 \pm 0.01$, and $0.09 \pm 0.06$, for L1450, L3000, and L5100, respectively.
In contrast, for the dust-obscured AGNs,
the $L_{\rm bol}$ values based on L1450, L3000, and L5100 tend to be lower than
those from $L_{\rm MIR}$, with $\log [L_{\rm bol}({\rm UV,\,opt})$/$L_{\rm bol}$(${\rm MIR}$)]=
$-0.46 \pm 0.06$, $-0.46 \pm 0.02$, and $-0.12 \pm 0.09$, respectively.

For dust-obscured AGNs, we applied extinction correction to the L3000 values,
which is insensitive to SED shapes and not limited by small number statistics.
By deriving $L_{\rm bol}$ values from the extinction-corrected L3000 values using the $E(B-V)$ measurements,
we found that the resulting $L_{\rm bol}$ values are consistent with the $L_{\rm MIR}$-based $L_{\rm bol}$ values,
yielding $\log [L_{\rm bol}({\rm 3000\,\AA{}})$/$L_{\rm bol}$(${\rm MIR}$)]=$0.03\pm0.02$.
These findings suggest that the underestimation of the L3000-based $L_{\rm bol}$ for dust-obscured AGNs arises from dust extinction.
A more conclusive verification of this result would requires accurate $E(B-V)$ measurements,
achievable through complementary analyses such as line ratio, X-ray hardness, and other diagnostics (e.g., \citealt{kim18a}).
However, such analyses are beyond the scope of this study.

Therefore, we concluded that 
the substantial fraction (34\,\%) of dust-obscured AGNs among the Type-1 AGNs in the NEP-Wide field
raises concerns that $L_{\rm bol}$ estimates may be significantly underestimated
if UV continuum luminosities are used as $L_{\rm bol}$ estimators without dust extinction correction.
Given the prevalence of dust-obscured AGNs across the entire AGN population, not just within the NEP-Wide field, 
our findings underscore that correcting for dust extinction is a crucial step in AGN studies. 

\subsection{Eddington Ratios of Dust-obscured AGNs} \label{sec:Edd}

In this subsection, we investigated whether dust-obscured AGNs exhibit higher $\lambda_{\rm Edd}$ values than unobscured AGNs.
In the merger-driven galaxy evolution scenario \citep{hopkins08},
dust-obscured AGNs are expected to have higher $\lambda_{\rm Edd}$ values than unobscured AGNs (\citealt{urrutia12,kim15a,kim24a,kim24b}),
and the extinction of dust-obscured AGNs arises from dust in their host galaxies.
Alternatively, if the extinction of dust-obscured AGNs originates from the dust torus viewed at moderate inclinations (\citealt{wilkes02,rose13})
in the orientation-based unification model,
dust-obscured AGNs are expected to have similar $\lambda_{\rm Edd}$ values to unobscured AGNs.

Moreover, to examine the dependence of $\lambda_{\rm Edd}$ on $E(B-V)$,
we divided the dust-obscured AGNs into two subsamples:
moderately obscured ($0.1 < E(B-V) \leq 0.2$) and heavily obscured ($E(B-V) > 0.2$),
following the classification in Section \ref{sec:SED_shape},
and compared their $\lambda_{\rm Edd}$ values with those of unobscured AGNs.
Note that in this comparison, only the AGNs at $z < 2.5$ were considered,
where the number of dust-obscured AGNs is sufficient to yield statistically robust results.

For this comparison, we used 296 unobscured AGNs and 125 dust-obscured AGNs. 
Among the dust-obscured AGNs, 77 objects are the moderately obscured AGNs,
and the remaining 48 objects are the heavily obscured AGNs.

The median $\log \left( {\lambda_{\rm Edd}} \right)$ values of the unobscured and dust-obscured AGNs
are $-1.23 \pm 0.02$ and $-1.16 \pm 0.04$, respectively, over the entire redshift range. 
This result shows that the dust-obscured AGNs have slightly higher $\lambda_{\rm Edd}$ values than the unobscured AGNs.
To quantify the significance of this difference,
we performed a Kolmogorov–Smirnov (K-S) test using the \texttt{SciPy} \citep{virtanen20} Python package.
The maximum deviation between the cumulative distributions, $\it D$, is 0.13,
and the corresponding $\it p$-value is 0.08,
which suggests that the two distributions differ at a marginally significant level.

Note that the median $\log \left( {\lambda_{\rm Edd}} \right)$ values of
the moderately obscured and heavily obscured AGNs are $-1.18 \pm 0.04$ and $-1.13 \pm 0.06$, respectively.
Moreover, we performed additional K-S tests comparing the two dust-obscured AGN subsamples with the unobscured AGNs.
The resulting $\it D$ values are 0.11 and 0.20, and the corresponding $\it p$-values are 0.43 and 0.05, respectively.
These results suggest that
(i) the $\lambda_{\rm Edd}$ values of dust-obscured AGNs are higher than those of the unobscured AGNs,
and (ii) the difference increases with $E(B-V)$, although this trend is not statistically robust. 

However, these $\lambda_{\rm Edd}$ comparisons can be affected by (i) sample bias or (ii) cosmological evolution. 
Therefore, we compared the $\lambda_{\rm Edd}$ values of the unobscured and dust-obscured AGNs as a function of redshifts.
For this purpose, we divided the two AGN populations into eight redshift bins: 
0.1--0.4, 0.4--0.7, 0.7--1.0, 1.0--1.3, 1.3--1.6, 1.6--1.9, 1.9--2.2, and 2.2--2.5.
The median $\log \left( {\lambda_{\rm Edd}} \right)$ values of the unobscured AGNs in these redshift bins are
$-1.61\pm0.41$, $-1.62\pm0.44$, $-1.29\pm0.35$, $-1.33\pm0.37$, $-1.09\pm0.31$, $-1.06\pm0.33$, $-0.94\pm0.33$, and $-1.03\pm0.35$,
whereas those of the dust-obscured AGNs are
$-1.43\pm0.35$, $-1.23\pm0.31$, $-1.31\pm0.42$, $-1.09\pm0.35$, $-1.21\pm0.33$, $-1.21\pm0.40$, $-0.92\pm0.32$, and $-0.94\pm0.32$.
These results suggest that although both the unobscured and the dust-obscured AGNs exhibit
an increasing trend in $\lambda_{\rm Edd}$ with redshift,
the difference in $\lambda_{\rm Edd}$ between the two populations is not statistically strong. 

Furthermore, the redshift dependence of $\lambda_{\rm Edd}$ was investigated for the two dust-obscured AGN subsamples.
The median $\log \left( {\lambda_{\rm Edd}} \right)$ values of the moderately obscured AGNs in the redshift bins are
$-1.79\pm0.00$, $-1.50\pm0.28$, $-1.31\pm0.31$, $-1.07\pm0.32$, $-1.32\pm0.23$, $-1.34\pm0.38$, $-0.92\pm0.25$, and $-0.94\pm0.35$,
while those of the heavily obscured AGNs are
$-1.29\pm0.35$, $-1.20\pm0.26$, $-1.39\pm0.54$, $-1.35\pm0.41$, $-0.96\pm0.49$, $-1.06\pm0.40$, $-0.64\pm0.39$, and $-0.81\pm0.23$.
These results show that the $\lambda_{\rm Edd}$ values of the moderately obscured AGNs are comparable to those of the unobscured AGNs.
In contrast, the heavily obscured AGNs exhibit somewhat higher $\lambda_{\rm Edd}$ values,
although the difference is not statistically robust.

\begin{figure}
\centering
\figurenum{10}
\includegraphics[width=\columnwidth]{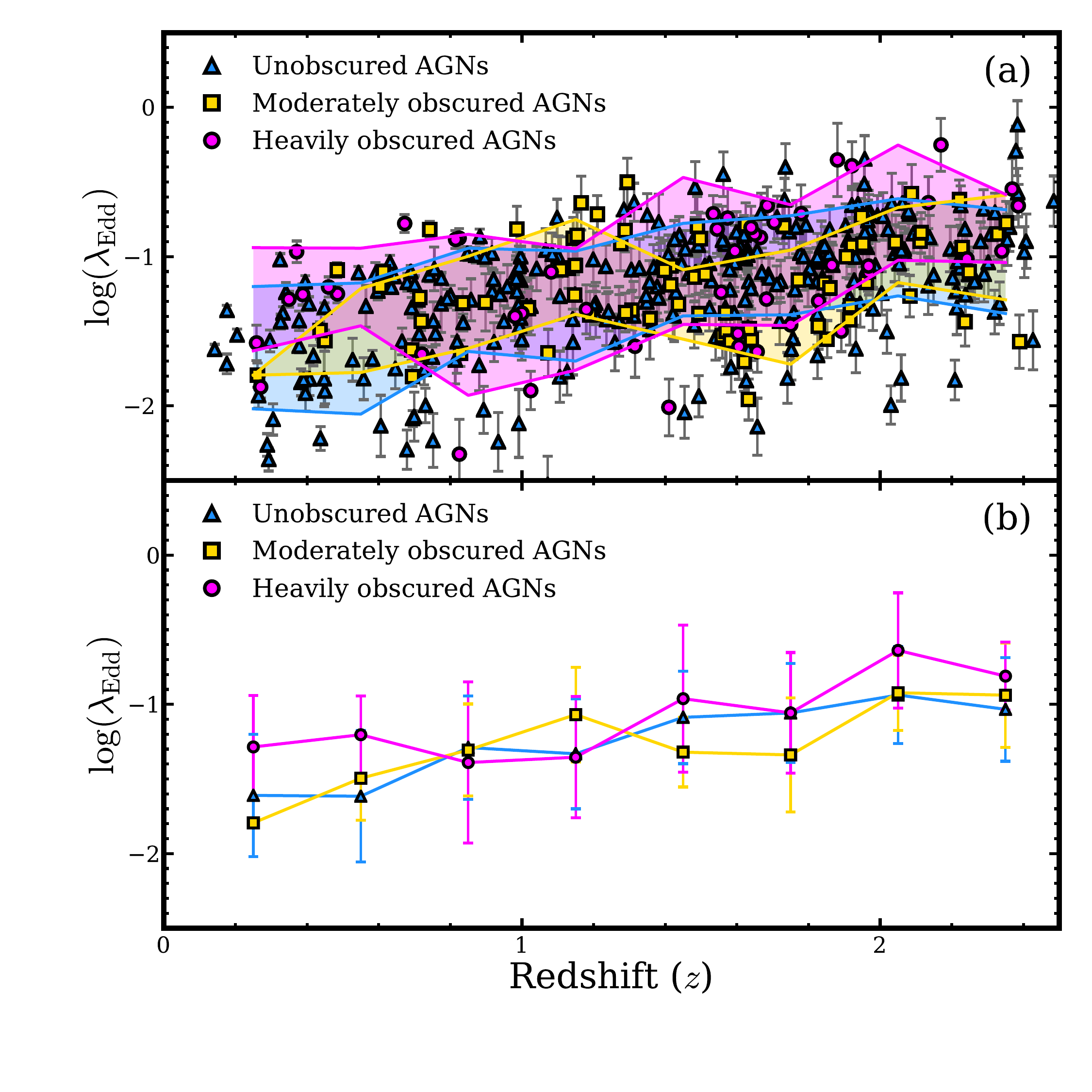}
\caption{
$\log \left( {\lambda_{\rm Edd}} \right)$ values versus redshifts.
Blue triangles, yellow squares, and magenta circles correspond to
unobscured ($E(B-V) \leq 0.1$), moderately obscured ($0.1 < E(B-V) \leq 0.2$), and heavily obscured ($E(B-V) > 0.2$) AGNs, respectively.
(a) Colored polygons indicate the median $\log \left( {\lambda_{\rm Edd}} \right)$ and 1$\sigma$ dispersions in eight redshift bins.
(b) Median $\log \left( {\lambda_{\rm Edd}} \right)$ values in the same eight redshift bins are shown with 1$\sigma$ dispersions.
\label{fig:RQ_Edd_z}}
\end{figure}

These comparisons are presented in Figure \ref{fig:RQ_Edd_z}.
We do not find statistically robust evidence for an increase in $\lambda_{\rm Edd}$ with $E(B-V)$,
consistent with an orientation-based interpretation in which
moderate obscuration arises from a dust torus viewed at intermediate inclination (e.g., \citealt{wilkes02,rose13}).
In contrast, the heavily obscured AGNs may exhibit higher $\lambda_{\rm Edd}$ values than the other populations,
consistent with previous studies (e.g., \citealt{urrutia12,kim15a,kim24a,kim24b}).
This suggest that their dust extinction may arise from dust in the host galaxies. 

\section{Summary} \label{sec:summary}

We derived the BH properties of 861 Type-1 AGNs in the NEP-Wide field 
using the $L_{\rm MIR}$ values at 3.4\,$\mu$m and 4.6\,$\mu$m,
combined with the FWHM values of the \ion{C}{4}, \ion{Mg}{2}, H$\beta$, and H$\alpha$ lines.
By performing SED and line fitting, we obtained the $L_{\rm MIR}$ and FWHM values.
From these measurements, we derived the $L_{\rm bol}$, $M_{\rm BH}$, and $\lambda_{\rm Edd}$ values,
which are less affected by dust extinction due to the use of the extinction-corrected $L_{\rm MIR}$ values.
We found that 34\,\% of the Type-1 AGNs in the NEP-Wide field are dust obscured,
which can lead to significant underestimation of their $L_{\rm bol}$ values
if UV luminosities are used as proxies for $L_{\rm bol}$ without applying proper extinction correction.
These findings underscore the importance of correcting for dust extinction in AGN studies.
As ongoing and upcoming IR spectroscopic missions, such as SPHEREx and $\it{Euclid}$, begin to cover the NEP-Wide field, 
our measurements will serve as fiducial BH properties for future AGN studies.
This paper represents the first step of our project, focusing on AGNs detected in both optical and MIR surveys.
In forthcoming studies (Papers II and III), we will extend our analysis to AGNs detected only in the optical survey and only in the MIR survey, respectively.

\begin{acknowledgments}
We thank the anonymous referee for the useful comments.
D.K. acknowledges the support by the National Research Foundation of Korea (NRF) grant 
(No. 2021R1C1C1013580 and 2022R1A4A3031306) funded by the Korean government (MSIT).
M.I. acknowledges the support from the National Research Foundation of Korea (NRF) grant,
No. 2021M3F7A1084525, funded by the Korea government (MSIT).
H.S. acknowledges the support from the National Research Foundation of Korea (NRF) grant funded
by the Korea government (MSIT) (No. RS-2024-00349364).
Y.K. was supported by the National Research Foundation of Korea (NRF) grant funded
by the Korean government (MSIT) (No. 2021R1C1C2091550).

We are grateful to Jie Chen for providing the linear relations between $L_{\rm bol}$ values and the UV/optical continuum luminosities,
which greatly contributed to this study.
\end{acknowledgments}

\clearpage

\end{document}